\documentclass[a4paper,11pt]{article}
\usepackage{graphicx}
\usepackage[space]{grffile}
\usepackage{latexsym}
\usepackage{textcomp}
\usepackage[normalem]{ulem}
\usepackage{colortbl}
\usepackage{amsfonts,amsmath,amssymb}
\usepackage{natbib}
\usepackage[hidelinks]{hyperref}
\usepackage{subcaption}
\usepackage{etoolbox}
\makeatletter
\usepackage{authblk}
\usepackage[utf8]{inputenc}
\usepackage[english]{babel}
\usepackage[top=0.8in, bottom=0.8in, left=0.8in, right=0.8in]{geometry}

\title{\textbf{The role of tropical and extra-tropical waves in the Hadley circulation}}

\author[1,2]{ABS Thakur  \thanks{Corresponding author: ABS Thakur, thakur.abubakar@gmail.com}}
\author[1,2]{Jai Sukhatme}
\author[3]{Nili Harnik}

\affil[1]{Centre for Atmospheric and Oceanic Sciences, Indian Institute of Science, Bangalore 560012, India}
\affil[2]{Divecha Centre for Climate Change, Indian Institute of Science, Bangalore 560012, India}
\affil[3]{Department of Geosciences, Tel Aviv University, Tel Aviv 69978, Israel}

\begin{document}
	
\maketitle
\begin{abstract}
    \normalsize{The tropical overturning circulation is examined in a moist aquaplanet general circulation model forced using a non-interactive sea surface temperature (SST) distribution that varies between a present-day Earth-like profile and one that is globally uniform. A traditional Hadley Cell (HC)-like flow is observed in all experiments along with the poleward transport of heat and angular momentum. In simulations with non-zero SST gradients, latent heat released from organized convection near the equator sets up a deep tropical cell; midlatitude baroclinic Rossby waves flux heat and angular momentum poleward, reinforcing the thermally direct circulation. As the imposed SST gradient is weakened, the HC transitions from a thermally and eddy-driven regime to one that's completely eddy-driven. When the SST is globally uniform, equatorial waves concentrate precipitation in the tropics and facilitate the lower-level convergence necessary for the ascending branch of the HC. Conventional midlatitude Rossby waves become very weak, but upper-level baroclinicity generates waves that cause equatorward transport of heat and poleward transport of momentum. Moreover, these upper-level waves induce a circulation that opposes the time-mean HC, thus highlighting the role of tropical waves in driving a traditional overturning flow for uniform SSTs. In all cases, anomalies associated with the tropical waves closely resemble those that sum to give the upper-level zonal mean divergent outflow. Through their ability to modulate tropical rainfall and the related latent heating, equatorial waves cause considerable hemispheric asymmetry in the HC and impart synoptic and intraseasonal variability to the tropical overturning circulation.} 
\end{abstract}

\section{Introduction}

The Hadley Cell (HC) is the most prominent feature of the zonally averaged tropical circulation. It consists of a pair of thermally direct cells with rising motion over the deep tropics and subsidence over the subtropical latitudes and transports energy and angular momentum from the equatorial region to the subtropics \citep{peixoto1992physics}. The presence of the HC explains the occurrence of easterlies in the tropics and partly explains the subtropical jet in the upper troposphere. In contrast, large-scale baroclinic eddies are at the core of the extratropical circulation theories and flux heat and momentum from the subtropics towards the poles. These eddies arise from the release of zonal mean available potential energy stored in the pole-equator temperature gradient of the Earth, created by the externally imposed differential heating from the sun \citep{pierrehumbert_baroclinic}. Convergence of angular momentum by these eddies drives the Ferrel cell and explains the midlatitude westerly jets and the surface westerlies \citep{schneider_review}. 

Moreover, baroclinic eddies have an influence on the strength and extent of the Hadley circulation \citep{schneider_review, held_review}. The importance of these midlatitude baroclinic eddies in shaping the Hadley circulation is best appreciated by suppressing them artificially \citep{schneider_lindzen, schneider1977, held_hou}. Such axisymmetric simulations yield meridional overturning circulations similar to the annual mean HC with a subtropical jet and surface trade winds. These studies show that the HC does not extend to the poles, even in the absence of eddies. However, the subtropical jet is too strong in these angular momentum-conserving flows. Further, \citet{davis2019} observed that the axisymmetric Hadley cell remains confined to the upper troposphere and extends down to the lower boundary only in the presence of significant viscous damping or eddy stresses. Thus, eddies modify the wind and temperature fields to give a more realistic tropical circulation. 

Another way to assess the impact of eddies is to compare axisymmetric and eddy-permitting simulations, identical to each other in every respect except in their ability to resolve eddies \citep[for example, ][]{williamsA, williamsB}. Using an idealized dry GCM, \citet{becker1997feedback} showed that the feedback of eddies onto the Hadley circulation is dependent on the amount of equatorial heating. \citet{kim2001hadley} showed that while only 38\% of the HC is directly attributable to eddy fluxes, an equal amount is due to changes in diabatic heating and surface friction induced by the eddies. \citet{walker2006eddy} used an idealized dry GCM to show that the Hadley circulation strength is largely related to the eddy momentum flux divergence at the latitude of the streamfunction maxima, over a wide range of flow parameters. Moreover, transitions of the HC between equinoctial and solstitial regimes in response to seasonally varying thermal forcing are sharper in eddy-permitting simulations in comparison to axisymmetric simulations due to eddy-mean-flow interactions \citep{bordoni2010regime}. In both moist and dry simulations, the meridional extent of the HC is highly correlated with the latitude at which baroclinic eddies become deep enough to reach the upper troposphere \citep{korty-schneider, levine-schneider}. Moist simulations forced by fixed SST boundary conditions yield comparable strengths for the axisymmetric and eddy-permitting circulation \citep{satoh1995, singhkuang2016}; but with interactive SSTs, the eddy-permitting HC is 2-3 times stronger \citep{singhetal2017}.

From the discussion above, the influence of midlatitude baroclinic eddies on the HC is evident. However, the influence of tropical transient modes on the HC is not as well understood. In the zonal mean, regions of convergence and divergence associated with these waves tend to cancel each other, but the precipitation survives since it is a positive-only quantity \citep{horinouchi}. Convectively coupled equatorial waves tend to modify the structure of the precipitation field in the tropics \citep{Kiladis-rev}. As latent heating related to precipitation maintains the ascending branch of the overturning circulation, equatorial waves can influence the HC. In the summer and winter solstitial seasons, the overturning circulations that average to give the HC are confined to specific regions where both synoptic and intraseasonal activity are important \citep{hoskins2020, hoskins2021, hoskins2023}. In these regions, the maxima in Outgoing Longwave Radiation (OLR) variability and meridional wind have characteristic structures associated with westward-moving mixed Rossby-Gravity waves \citep{hoskins2020}. Additionally, \citet{hoskins2023} demonstrated that the upper-level circulation features associated with longitudinally localized deep convection can be described as the superposition of $n$=0 Mixed Rossby-Gravityand $n$=1 equatorial Rossby waves. Further, the local Hadley and Walker circulations are strengthened or weakened, based on the location of Madden-Julian Oscillation (MJO)-related active or suppressed convection \citep{schwendike}. In fact, equatorial waves contribute directly to the Atlantic HC strength through their meridional wind component and indirectly by triggering moist convection through low-level convergence \citep{tomassini2022}.

In idealized aquaplanet conditions, the role of equatorial waves is best illustrated in simulations with globally uniform sea-surface temperatures (SSTs) and solar insolation in the absence of any externally imposed meridional gradients except for the meridionally changing effect of the Earth's rotation \citep{sumi}. Well-defined structures like the intertropical convergence zone (ITCZ) and Hadley circulation emerge in such simulations \citep{sumi, kirtman, chao2004single, barsugli, horinouchi, suhas2021}. Here, \citet{sumi} observed that the organization of convection was sensitive to the rotation rate of the Earth and reported eastward propagating convective activity in the tropics after the setting-up of zonal mean states. Varying a tunable parameter that affects the tropical transient variability, \citet{horinouchi} obtained nearly symmetric HCs with strengths that are comparable to the observed equinoctial climatological values; moreover, the strength of these HCs were highly correlated with the eddy kinetic energy in the tropical lower troposphere. By modulating the latent heat of vaporization of water ($L_v$), \citet{suhas2021} mimicked atmospheres with varying degrees of moist coupling. They observed that the strength of their simulated HC decreases with weaker moist coupling; and there was a shift in the dominant intraseasonal peak, from 90 to 30 days, linked with eastward propagating modes associated with convergence and moist ascent over the equator. 

Thus, it is apparent that eddies of both tropical and extratropical origin can influence the tropical overturning circulation. So, a theory for the dynamics of the HC should incorporate both tropical and extratropical eddies. In this study, our primary motivation is to showcase the role of tropical waves in contrast to extratropical waves in shaping the Hadley circulation. We use an aquaplanet general circulation model forced by a range of zonally symmetric lower boundary forcing conditions along with globally uniform solar flux and fixed orbital parameters. Specifically, we vary SSTs gradually from present-day Earth-like conditions to one with a globally uniform latitudinal profile. Varying the SSTs in this manner gives us a passive control on the generation of midlatitude baroclinic eddies, as the growth rate of these instabilities is directly proportional to the meridional temperature gradient \citep{eady}. Therefore, we expect to transition from an Earth-like atmosphere with dominant midlatitude eddies to one where they are very weak or non-existent.

Apart from a fundamental insight into the role of eddies in the HC, the varying SST simulations are relevant in the global warming scenario as a robust response of the climate system to greenhouse gas forcing is a decrease in pole-equator temperature difference \citep{vallis2015}. 
Although we do not change the Carbon Dioxide concentration in our simulations, by weakening the meridional gradient of SST we are forcing the indirect atmospheric response to greenhouse gas warming \citep{deser, polvani_ozone, shaw2015tug}. Also, as noted by \citet{sumi}, an atmosphere with globally uniform SST is possible when the planet's interior heating by long-wave radiation is dominant in comparison to external heating by short-wave radiation. Such an outcome is possible in a runaway greenhouse scenario \citep[for example, Venus;][]{ingersoll, goldblatt}. 
Moreover, relatively flatter SSTs were known to have prevailed in paleoclimates, thus there is relevance in understanding how the atmospheric circulation would have differed in such a setting \citep{kortyemanuel, fedorov2010tropical, fedorov2019tropical}. 

The rest of the paper is structured as follows: Section 2 presents details of the modeling framework, the experimental setup, along with the diagnostic tools used in this study. Section 3 presents the zonally averaged picture. The role of the midlatitude and tropical eddies in shaping the zonal mean circulation is discussed in Section 4. Section 5 contains a discussion of our findings along with conclusions and future scope.

\section{Methods and Analysis}

\subsection{Simulations}

\begin{figure}[h]
	\centering
	\includegraphics[width=0.7\linewidth]{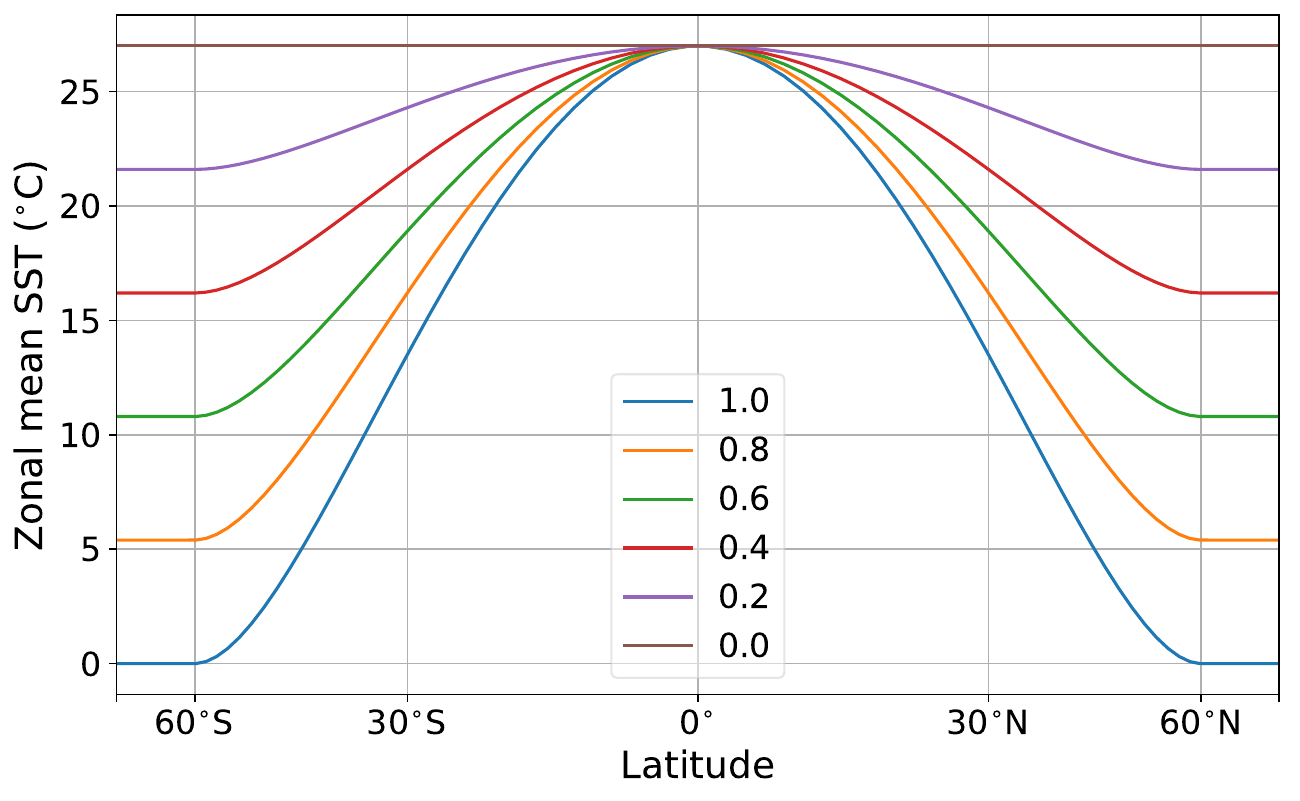}
	\caption{Zonal mean profiles of the sea surface temperatures (SST) for $\alpha \in [0,1]$ used to force the aquaplanet simulations. }
	\label{fig1}
\end{figure}

We carry out aquaplanet simulations using the Community Atmosphere Model (CAM6) of Community Earth System Model \citep[CESM v2.1.3;][]{danabasoglu2020community}. The simulations are performed using the Eulerian dynamical core which is a global spectral model, at T42 horizontal resolution and 32 vertical levels.  The model uses the Zhang-MacFarlane convection scheme for deep convection and the Cloud Layers Unified By Binormals (CLUBB) scheme for both shallow convection and microphysics. All simulations are performed with uniform solar insolation of 342 W-m$^{-2}$, perpetual equinox orbital condition, and no diurnal cycle. As we are primarily focused on the physical processes that govern the tropical overturning circulation, the chemistry model is turned off to reduce computational cost. Following \citet{medeiros2016reference}, we make the aerosols radiatively inactive. The model is forced by zonally symmetric sea surface temperature (SST) profiles shown in Figure \ref{fig1}. The SST profiles range from approximating the present-day climate (warm equator and cold poles, $\alpha$=1.0) to being uniform with latitude ($\alpha$=0.0). The SST profile for the $\alpha$=1.0 run is the \textit{Qobs} profile of \citet{neale}. The analytical expressions for the SST profiles are as follows.
\begin{equation}
        T_s = 
    \begin{cases}
    T_0 \left(1-\alpha\sin^2(\frac{3 \phi}{2}) \right),& \vert\phi\vert < \frac{\pi}{3} \\
    T_0 (1-\alpha),&         \text{otherwise}
    \end{cases} \\
\end{equation}

As discussed in the Introduction, by holding the tropical SST fixed and gradually warming the poles, we are emulating the weakening of the meridional surface temperature gradients and polar amplification which are robust trends in global warming simulations \citep{vallis2015}. Indeed, a similar reduction in SST gradients can be achieved by holding the polar SST fixed and cooling the tropics, slowly transitioning from the present-day Earth-like climate to a Snowball Earth \citep{pierrehumbert_snowball, voigt_snowball}.

Note that we do not relax to a radiative equilibrium profile. The cells and winds develop spontaneously as a response to boundary forcing. The impact of eddies on the Hadley circulation is analyzed using eddy-free axisymmetric simulations. In this set, all wave components ($k>0$) are set to zero \citep{satoh1994, becker1997feedback, bordoni2010regime}. The utility of this set will be explained in Section \ref{sec:KE_decomp}, and the corresponding results will be presented in Section 4. All simulations are run for 12 years, and the results presented in this study are based on the last 11 years.


\subsection{Zonal mean and eddy diagnostics}

Here we provide a brief description of the diagnostic utilities employed in this study to assess the zonal mean circulation. In all the expressions that follow, square brackets ($[.]$) indicate a zonal mean while the overbar $\overline{(.)}$ indicates a temporal mean. Primed quantities ()' refer to eddies.

\subsubsection{\normalsize{Mass-weighted streamfunction}}

The Eulerian mean mass-weighted streamfunction is defined as \citet{peixoto1992physics},

\begin{equation}
    \Psi = \frac{2\pi a\cos \phi}{g} \int_{0}^{p_s}[\overline{v}]dp,
    \label{eq:mwsf}
\end{equation}

where $a$ is the radius of the Earth, $g$ is the acceleration due to gravity and $p_s$ is the surface pressure. The above definition satisfies the zonal mean continuity equation such that $([\overline{v}], [\overline{\omega}])=(\frac{\partial \Psi}{\partial p}, -\frac{\partial \Psi}{a \partial \phi})$.

\subsubsection{\normalsize{Decomposing the eddy momentum flux into rotational and divergent components}}

Treating the horizontal wind field on each pressure level as a two-dimensional vector field, the rotational and divergent components can be obtained using the Helmholtz decomposition. Following \citet{zurita2019role}, the rotational (denoted by subscript $r$) and divergent (subscript $d$) components of the eddy momentum flux are,
\begin{align}
    u'v'|_r = u_r'v_r', \\ \nonumber
    u'v'|_d = u_r'v_d'+u_d'v_r'.
\end{align}
The contribution from the purely divergent component $u_d'v_d'$ is much smaller compared to the other components \citep{zurita2019role}.  

\subsubsection{\normalsize{Rossby number}}

We make use of the vorticity-based bulk Rossby number formulation, as defined in \citet{martin-notes}.
$Ro$ is computed at the latitude of the streamfunction maxima as 

\begin{equation}
    Ro=-\frac{1}{f}\frac{\int_{p_t}^{p_m}{[\overline{v}][\overline{\zeta}]\frac{dp}{g}}}{\int_{p_t}^{p_m}{[\overline{v}]\frac{dp}{g}}},
    \label{eq:rossby_num}
\end{equation}
where $p_m$ is the level of the streamfunction maxima and $p_t$ is the tropopause level, $f$ is the Coriolis parameter and $v$ is the meridional wind.

\subsubsection{\normalsize{Vertically integrated zonal mean transport of energy and momentum}}

The poleward transport of moist static energy (MSE) is, $$2 \pi a \cos{\phi}\int_{0}^{p_s}{\overline{v MSE}} \frac{dp}{g},$$

where MSE consists of the dry static energy (DSE) and latent energy. The poleward transport of momentum is defined as in \citet{satoh1995}, $$2 \pi a^2 \cos^2{\phi}\int_{0}^{p_s}{\overline{uv}} \frac{dp}{g}.$$

\subsubsection{\normalsize{Eliassen-Palm flux}}

The generation, propagation and dissipation characteristics of Rossby wave activity are explored through the Eliassen-Palm (EP) flux. As in \citet{edmon1980eliassen}, the zonal mean EP flux ($\boldsymbol{F}$) is, 

\begin{equation}
    \boldsymbol F \equiv \{F_\phi, F_p\} = a \cos{\phi} \biggl\{-[\overline{u'v'}], f\frac{[\overline{v'\theta'}]}{\partial_p\overline{[\theta}]}\biggr\}.
    \label{eq:epflux}
\end{equation}
Here, primed quantities denote deviations from a zonal mean. When WKBJ theory is valid, $\boldsymbol{F}$ ($ \approx \boldsymbol{c_g}\mathcal{A}$) is the advective flux of wave activity ($\mathcal{A}$) by the group velocity ($\boldsymbol{c_g}$). For frictionless and adiabatic flow, the wave activity follows the conservation relation \citep{edmon1980eliassen}, $$\frac{\partial \mathcal{A}}{\partial t} + \boldsymbol{\nabla \cdot F} = 0,$$ where,

\begin{equation}
    \boldsymbol{\nabla \!\cdot F} = \frac{1}{a \cos{\phi}}\frac{\partial }{\partial \phi}(F_{\phi}\cos{\phi}) + \frac{\partial}{\partial p}(F_p) = (a \cos{\phi}) \overline{v'q'}.
\end{equation}

\subsubsection{\normalsize{Residual Circulation}}

As defined in \citet{edmon1980eliassen}, the residual velocity is, $$[\overline{v^*}] = [\overline{v}] - \partial_p \biggl( \frac{[\overline{v'\theta'}]}{\partial_p\overline{[\theta}]} \biggr).$$
Therefore, the residual mean meridional circulation streamfunction is given by,
\begin{equation}
    \Psi^* = \frac{2\pi a\cos \phi}{g} \int_{0}^{p_s}[\overline{v^*}]dp.
    \label{eq:tem}
\end{equation}

\subsection{Estimating the eddy influence on the overturning circulation\label{sec:KE_decomp}}

To study the impact of eddies on tropical overturning circulation, we utilize the Kuo-Eliassen equation \citep[][]{kuo1956forced, chemke2019opposite, kang2019tropical, zaplotnik}. This is a second-order linear elliptic partial differential equation that is generally derived using the quasi-geostrophic approximation \citep{peixoto1992physics}. Specifically, the quasi-geostrophic approximation is applied to the zonally averaged thermodynamic and momentum equations written in flux form. Using the zonally averaged continuity equation, a streamfunction ($\psi$) is introduced as $([v],[\omega])=\biggl(\frac{g}{2 \pi a \cos{\phi}}\frac{\partial \psi}{\partial p}, -\frac{g}{2 \pi a \cos{\phi}}\frac{\partial \psi}{a \partial \phi} \biggr)$ to represent the zonal mean non-divergent flow in the pressure-latitude plane. The zonal mean thermal wind relation ($f\frac{\partial [u]}{\partial p}=\frac{R}{p} \frac{\partial [T]}{\partial y}$) is used to combine the momentum and thermodynamic equations into the Kuo-Eliassen equation, which reads,

\begin{equation}
    \begin{aligned}
    f^2 \frac{g}{2 \pi a \cos{\phi}} \frac{\partial^2 \psi}{\partial p^2} + \Gamma \frac{g}{2 \pi a} \frac{\partial}{a \partial \phi} \biggl(\frac{1}{a \cos{\phi}} \frac{\partial \psi}{\partial \phi} \biggr) = \underbrace{\frac{R}{p} \frac{\partial [Q]}{a \partial \phi}}_{S_{Q}} -  \underbrace{f  \frac{\partial [D]}{\partial p}}_{S_F} \\- \underbrace{\frac{R}{p} \frac{\partial}{a \partial \phi} \frac{1}{a \cos{\phi}}\frac{\partial }{ \partial \phi} [v'T'] \cos{\phi} }_{S_{ehf}} - \underbrace{f  \frac{1}{a \cos^2 \phi}\frac{\partial^2}{ \partial p \partial \phi} [u'v']\cos^2 \phi }_{S_{emf}}. 
    \end{aligned}
    \label{eq:kuo-eliassen}
\end{equation}
In the above equation, $\Gamma$ ($ = -[T]\frac{\partial [\theta]}{\partial p}$) is the static stability parameter, $Q$ is the diabatic heating rate estimated from the thermodynamic budget, $D$ is the drag estimated from the momentum budget, $f$ is the Coriolis parameter, $R$ is the universal gas constant and $a$ is the radius of the Earth. Subscripts \textit{ehf} and \textit{emf} stand for eddy heat flux and eddy momentum flux, respectively.

It is convenient to write Equation \ref{eq:kuo-eliassen} in symbolic form as $$L \psi= S,$$ where $L$ is the linear operator on the LHS, and $S$ ($= S_{Q} + S_{ehf} + S_{F} + S_{emf}$) is the sum of the source terms on the RHS, and $\psi$ is the streamfunction. As the Kuo-Eliassen equation is linear, it allows for the diagnosis of the effect of various heat and momentum forcing protocols on the meridional overturning circulation. We solve the above equation for $\psi$ numerically by first converting Equation \ref{eq:kuo-eliassen} into a finite-difference equation on a pressure-latitude grid that is solved iteratively using Gauss-Seidel's method for faster convergence.

On long timescales, the source terms in Equation \ref{eq:kuo-eliassen} tend to affect each other, wherein the eddy fluxes will modify the diabatic heating and drag terms \citep{chang1996}. We utilize the methodology outlined by \citet{kim2001hadley} to estimate the true impact of the eddies on the Hadley circulation. This approach decomposes each term in the Kuo-Eliassen equation into an axisymmetric component (subscript \textit{a}) and the corresponding deviation due to the presence of eddies (subscript \textit{e}). Thus, $L=L_a+L_e$, $S=S_a+S_e$ and $\psi=\psi_a+\psi_e$. Noting that $L_a \psi_a = S_a$, we get \textit{purely eddy-driven} streamfunction response as \citep{kim2001hadley},
\begin{equation}
    \psi_e = L^{-1}(S_e-L_e\psi_a).
    \label{eq:kim-lee}
\end{equation}
This approach requires explicit knowledge of both axisymmetric and eddy-permitting simulations. It allows for the partitioning of the eddy influence on the overturning circulation into direct and indirect impacts. The direct impact of the eddies is via their heat and momentum transport. The indirect effect is through a modification of the diabatic heating, zonal mean temperature, static stability, and zonal mean zonal winds; this results in a change both in the diabatic forcing ($S_F$) and drag ($S_D$), and in the operator $L$, which depends on $\Gamma$. The effect of these changes on the Hadley circulation is the indirect impact of the eddies. Simply using the eddy-permitting simulation to diagnose the impact of eddies will result in the incorrect attribution of the indirect response of the eddies to the physical processes like drag or diabatic heating \citep{kim2001hadley}.

We have also performed this decomposition using the extended form of the Kuo-Eliassen equation \citep{zaplotnik}, which does not employ the quasi-geostrophic approximation in the momentum and thermodynamic equations before connecting them using thermal wind. The results are largely similar to the conventional form of the Kuo-Eliassen equation used in this study (Equation \ref{eq:kuo-eliassen}).

\section{The Zonal Mean View of the Circulation}

\begin{figure}[t]
	\centering
	\includegraphics[width=\linewidth]{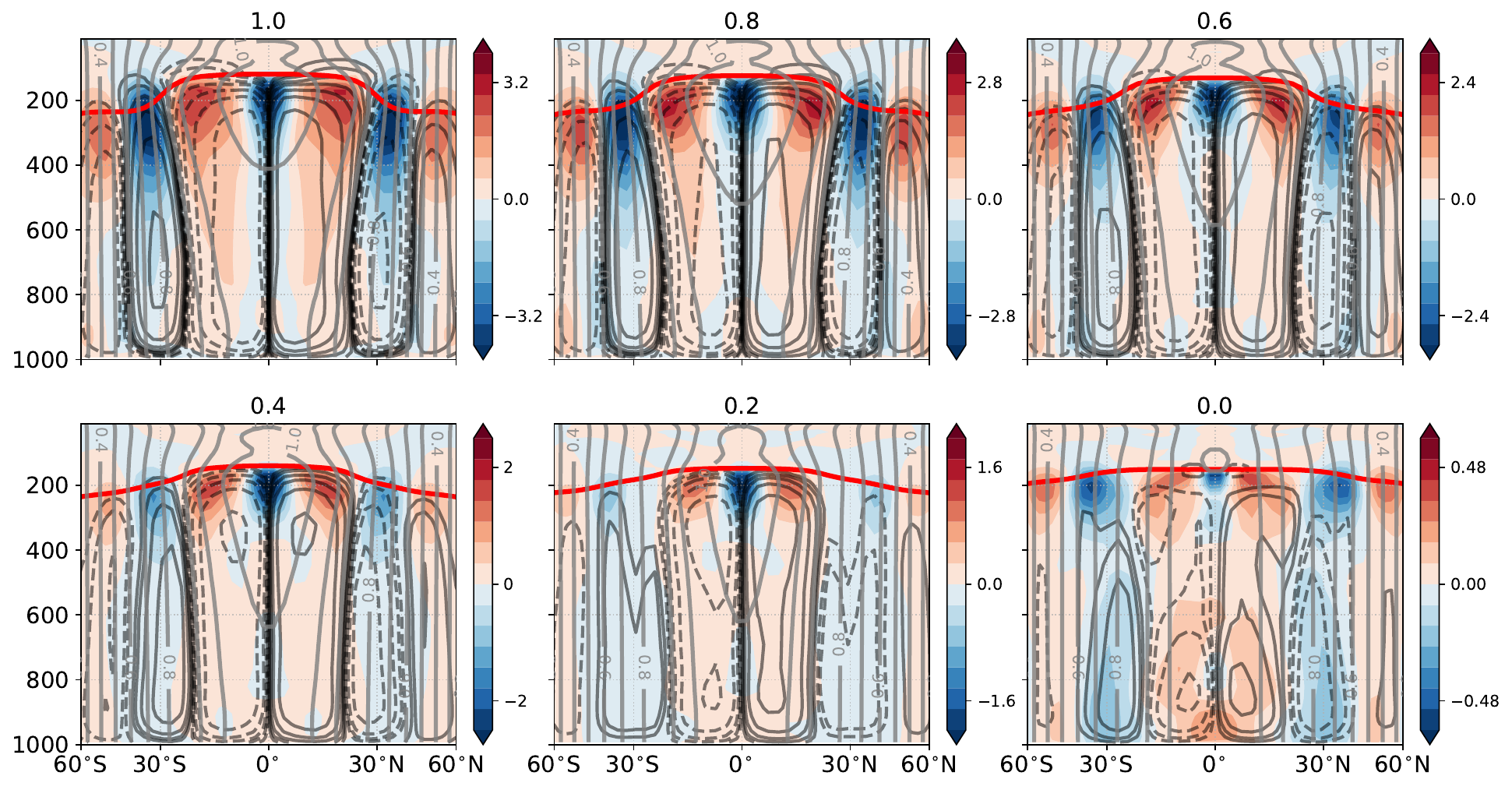}
	\caption{Eulerian mean mass streamfunction (dark contours) estimated using Equation \ref{eq:mwsf}, zonal mean horizontal eddy momentum flux divergence (colours) and angular momentum (light contours) for $\alpha \in [0,1]$. The eddy momentum flux divergence is presented in units of m s$^{-1}$day$^{-1}$. Streamfunction contours are logarithmic, with the magnitude of the lowest contour being 10$^{9}$ kg-s$^{-1}$. The magnitude doubles for successive contours. Contours of angular momentum are spaced at $0.1 \Omega a^2$ with the $0.95 \Omega a^2$ contour plotted additionally. The red curve in each panel shows the tropopause height \citep{reichler2003determining}.
	}
	\label{fig:psi_emfd_m}
\end{figure}

\subsection{Tropical overturning flow}

The Eulerian mean mass-weighted streamfunction is shown in Figure \ref{fig:psi_emfd_m} with black contours. The HC for the $\alpha$=1.0 simulation is similar to that obtained by other moist aquaplanet simulations \citep{satoh1995, williamson2013aqua, singhkuang2016, medeiros2016reference}, and the present-day Earth during the equinoctial seasons \citep{dima2005tropical}. For all $\alpha$, the temporally and zonally averaged tropical overturning flow consists of rising motion over the equator and descent in the subtropics. In all simulations, the overturning mean flow is accompanied by a poleward flux of DSE and an equatorward flux of latent energy; the tropical MSE transport is determined by the compensation between these components (Figure S1). As the equator-to-pole SST gradient decreases (i.e., smaller $\alpha$), we observe that the HC tends to be more localized towards the equator. In all cases, we see a well-formed Ferrel Cell poleward of the tropical HC. The intensity of the tropical circulation decreases with $\alpha$, and the maximum value of the streamfunction reduces by an order of magnitude when $\alpha$ goes to zero. 

\begin{figure}[t]
\centering
\captionsetup{justification=centering}
\begin{subfigure}{.55\textwidth}
  \centering
  \includegraphics[width=\linewidth]{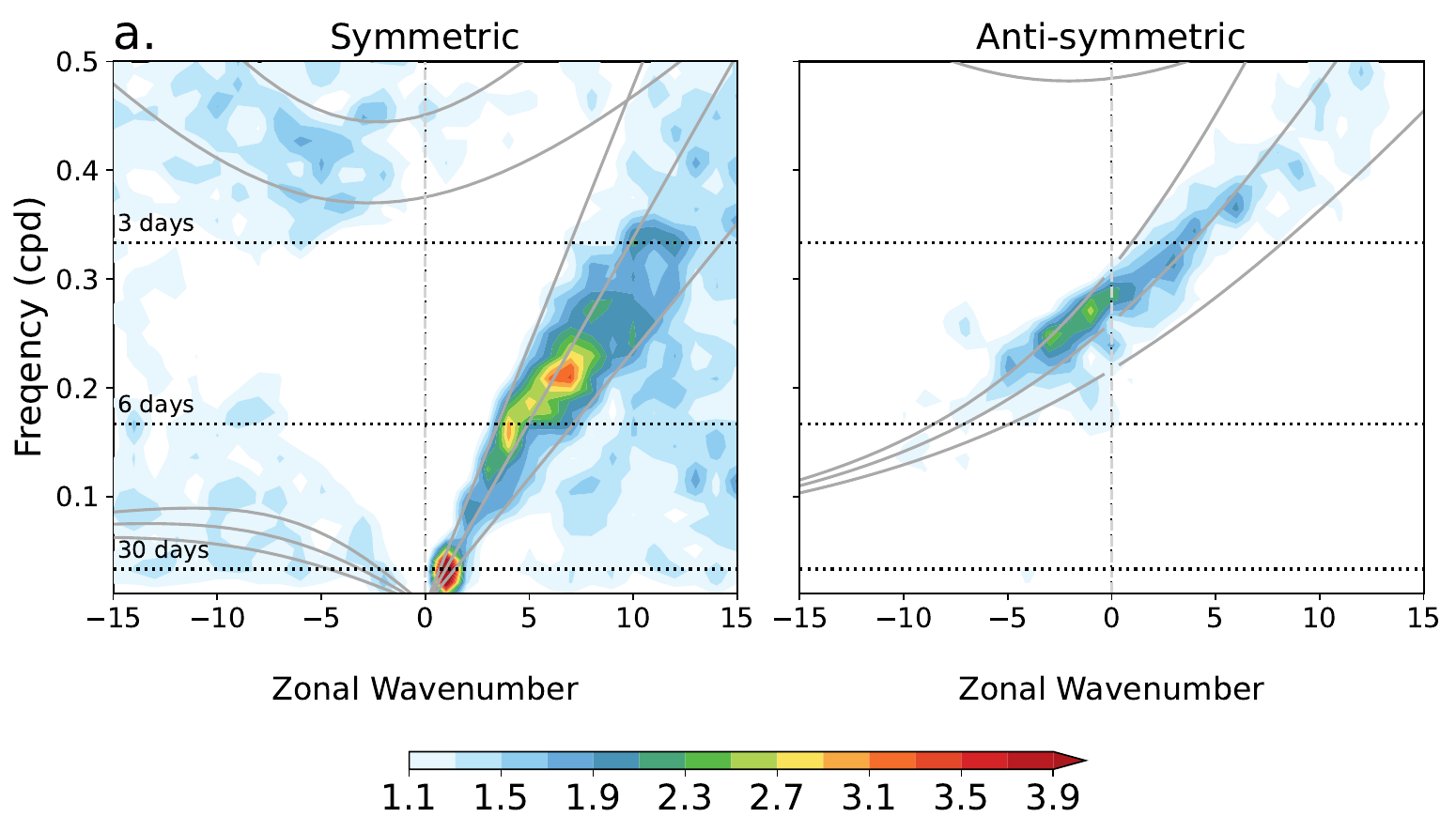}
\end{subfigure}%
\begin{subfigure}{.45\textwidth}
  \centering
  \includegraphics[width=\linewidth]{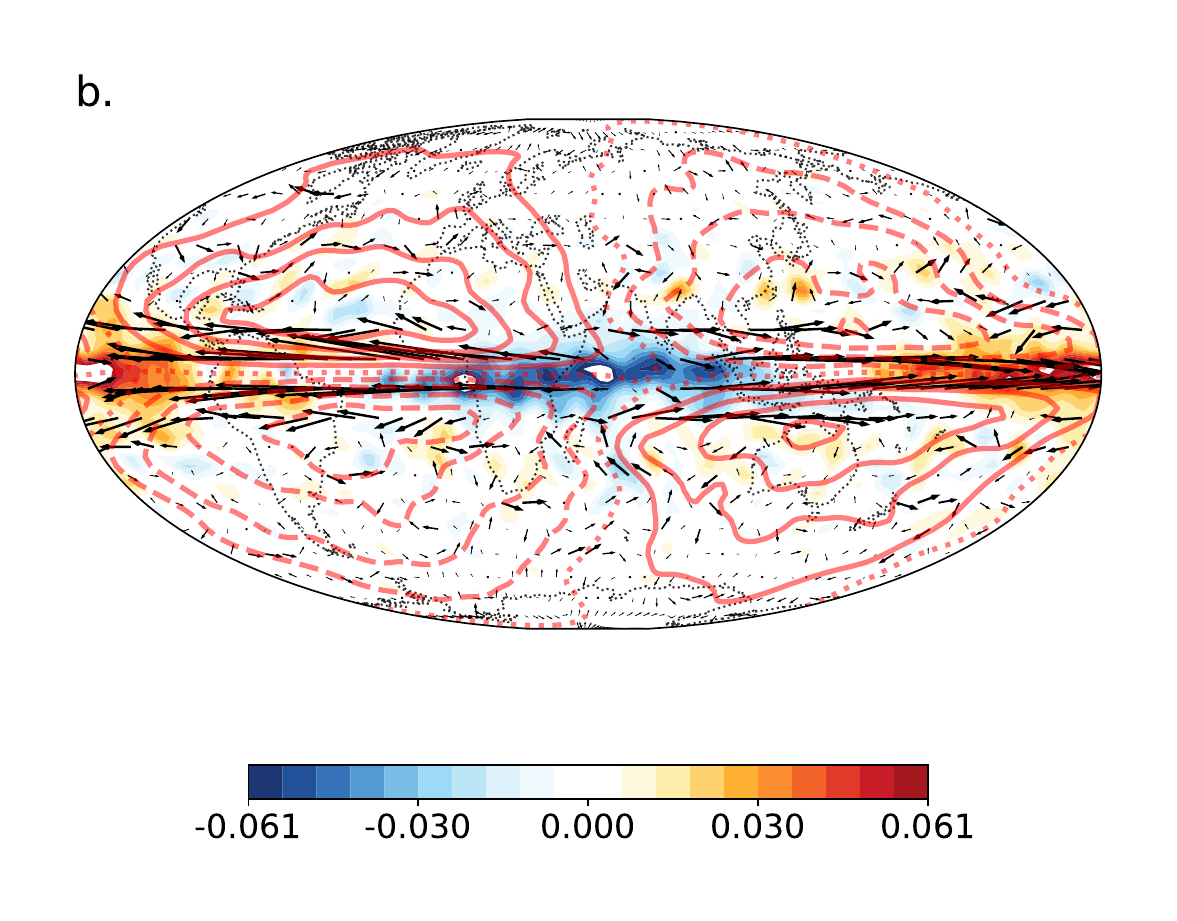}\par\medskip
\end{subfigure}
\begin{subfigure}{.5\textwidth}
  \centering
  \includegraphics[width=\linewidth]{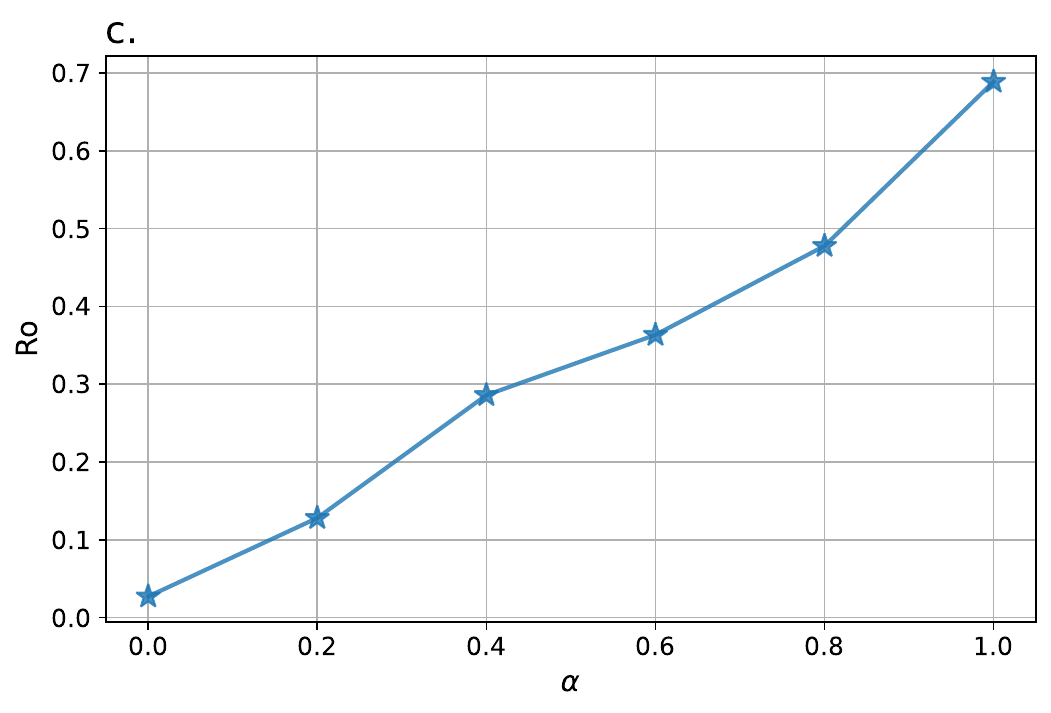}\par\medskip
\end{subfigure}
    \caption{For the globally uniform SST case. (a) Wavenumber-frequency power spectra of the vertical pressure velocity ($\omega$) at 850 mbar, calculated following \citet{wheeler1999convectively}. (b) Composite map of the Madden-Julian Oscillation \cite[MJO, ][]{zhang2005madden}. Colours represent $\omega$ at 500 mbar, while the contours and quivers represent the streamfunction and horizontal wind velocity averaged over the 150-300 mbar layer. Composites are obtained by regression onto the first two principal components of the 850 minus 150 mbar $\Delta \chi$ field, as outlined by \citet{aw1}. Continental boundaries are drawn in panel (b) for reference. (c) Vorticity-based bulk Rossby number (Equation \ref{eq:rossby_num}) comparison for all values of $\alpha$.}
\label{fig:wk_mjo_ro}
\end{figure}

In the cases with a non-zero SST gradient, deep convection preferentially organizes over the near-equatorial regions that have warmer SSTs; the circulation is then sustained by the equatorward flux of moisture by the mean flow (Figure S1). We also obtain an HC with globally uniform SSTs. Its strength is comparable to that observed in previous studies \citep{kirtman, barsugli, horinouchi, shi2014large, suhas2021}; though it is relatively weaker in the upper troposphere. A wavenumber-frequency diagram of tropical variability shows the presence of an eastward-moving Kelvin wave, a westward Mixed Rossby-Gravity wave along with a prominent low-frequency MJO-like mode \citep[Figure \ref{fig:wk_mjo_ro}a,b; see also][]{pritchard2016mjo, suhas2021}\footnote{We refer to this large-scale, eastward propagating system as an MJO-like mode as it has certain features in common with the Madden-Julian Oscillation, but there are differences as have been noted in other aquaplanet experiments \citep{das2016low,shi2018wishe}.}. These waves tend to concentrate the precipitation in the tropics, whereafter convection and circulation interactively adjust to set up a conventional HC \citep{horinouchi}.

Neglecting the vertical mean advection and eddy terms \citep{walker2006eddy, geen}, the temporally and zonally averaged zonal momentum balance in the upper troposphere can be written as $$f(1-Ro)[\overline{v}]=(\cos^2 \phi)^{-1}\partial_y[\overline{u'v'}\cos^2{\phi}].$$ 
Here, $Ro$ is the vorticity-based bulk Rossby number (Equation \ref{eq:rossby_num}), and other symbols have their usual meaning. The above equation implies a balance between zonal mean absolute vorticity flux by the mean flow (LHS) and the horizontal eddy momentum flux divergence (RHS). As $Ro \to 0$, the circulation responds to the eddy momentum flux divergence in the upper troposphere. As $Ro \to 1$, the circulation gets uncoupled from the eddy fluxes and responds directly to the thermal driving. In our experiments, $Ro$ decreases monotonically from about $0.689$ for $\alpha$=1.0 to $0.027$ for $\alpha$=0.0 (Figure \ref{fig:wk_mjo_ro}c). This implies that with decreasing $\alpha$, the HC transitions from a regime that is both thermally and eddy-driven to one that is strongly influenced by eddies. The regime change with $\alpha$ can be identified from the angular momentum contours as well. For the larger $\alpha$ runs, contours of constant angular momentum appear to be advected by the upper tropospheric flow. As seen in Figure \ref{fig:psi_emfd_m}, streamlines cross angular momentum contours towards the edge of the HC due to angular momentum transport out of the HC by eddies. As the HC moves closer to the low Rossby number regime with weakening SST gradients, angular momentum contours become more vertical, similar to those in the extratropics.

\begin{figure}[t]
	\centering
	\includegraphics[width=\linewidth]{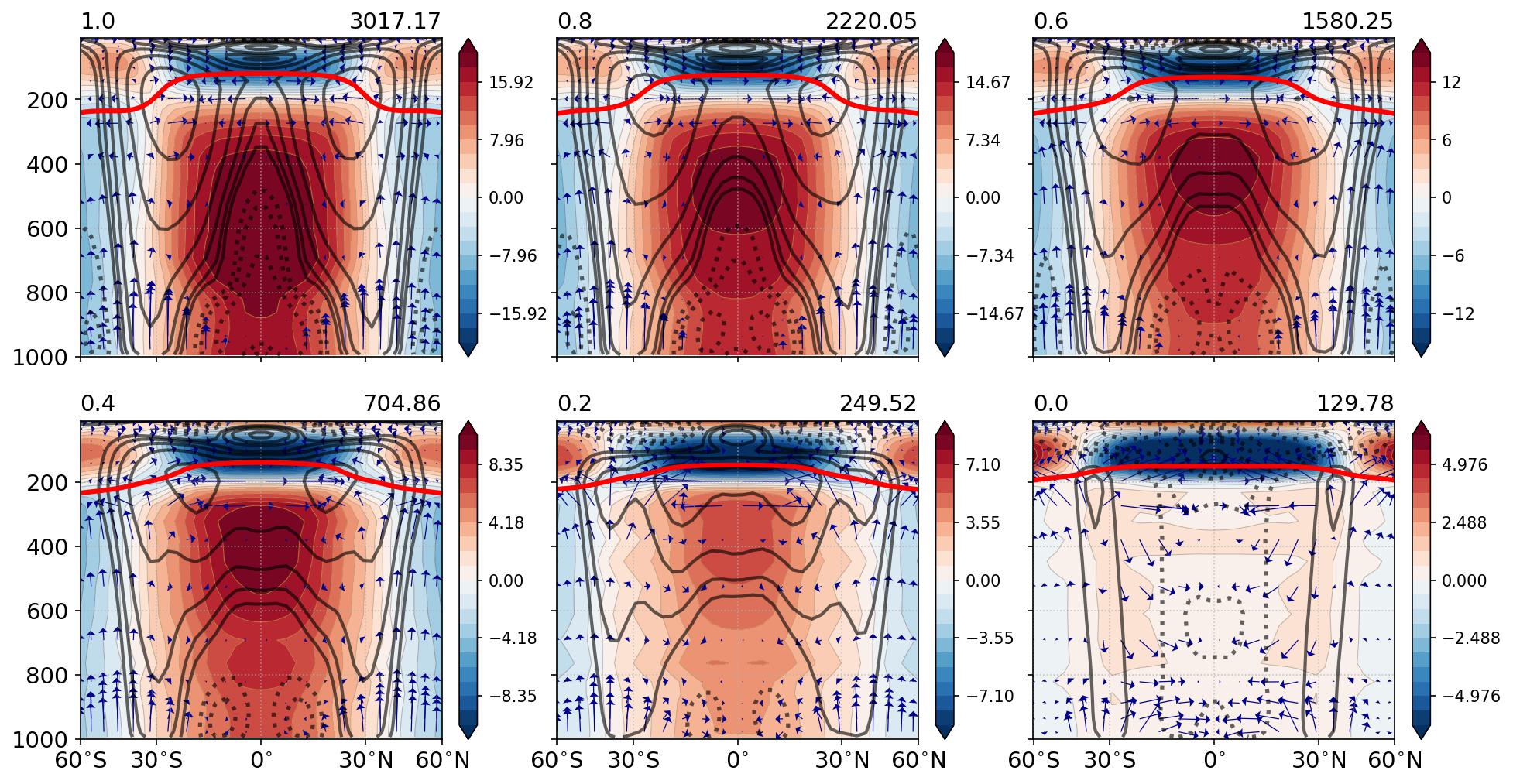}
	\caption{Anomalous temperature (colours), zonal mean zonal wind (contours) and Eliassen-Palm flux (Equation \ref{eq:epflux}; arrows) for $\alpha \in [0,1]$. Temperature anomalies are computed by removing the global mean from every pressure level. Zonal wind contours are logarithmic in nature. The smallest magnitude represented is $1$ ms$^{-1}$, and the magnitude doubles for successive contours. The number of data units per unit length of the arrows is shown on the top-right of each panel. The red curve in each panel shows the tropopause height \citep{reichler2003determining}. Prior to the presentation, the EP flux vectors were scaled as $\{ \frac{k \cos{\phi}}{\pi} F_{\phi}, \frac{k}{10^5} F_p \}$ where $k=\cos{\phi}\sqrt{\frac{1000}{p}}$}.
	\label{fig:u_ep_t}
\end{figure}

\subsection{Eddy momentum and EP flux}\label{sec:emf_epf}

The colours in Figure \ref{fig:psi_emfd_m} show the divergence of the horizontal eddy momentum flux (EMF). For pronounced SST gradients, eddy momentum tends to maximize at the tropopause because typical baroclinic eddies reach that level and break near the subtropical edge of the HC \citep{ait2015eddy}. The EP fluxes for the $\alpha$=1.0 run (arrows in the top-left panel of Figure \ref{fig:u_ep_t}) resemble the present-day atmosphere; wave activity moves upwards in the midlatitudes, then turns equatorward near the tropopause which acts as a turning surface for baroclinic Rossby waves \citep{ait2015eddy}. A rotational-divergent partition of the EMF (Figure S2) reveals that this is due to the rotational component, typically associated with midlatitude eddies. The EMF convergence in the deep tropics is due to the divergent component, linked with the MJO-like mode \citep[see Figures \ref{fig:wk_mjo_ro}a,b, S2 and S7;][]{zurita2019role}.

\begin{figure}[t]
	\centering
	\captionsetup{justification=centering}
	\includegraphics[width=0.95\linewidth]{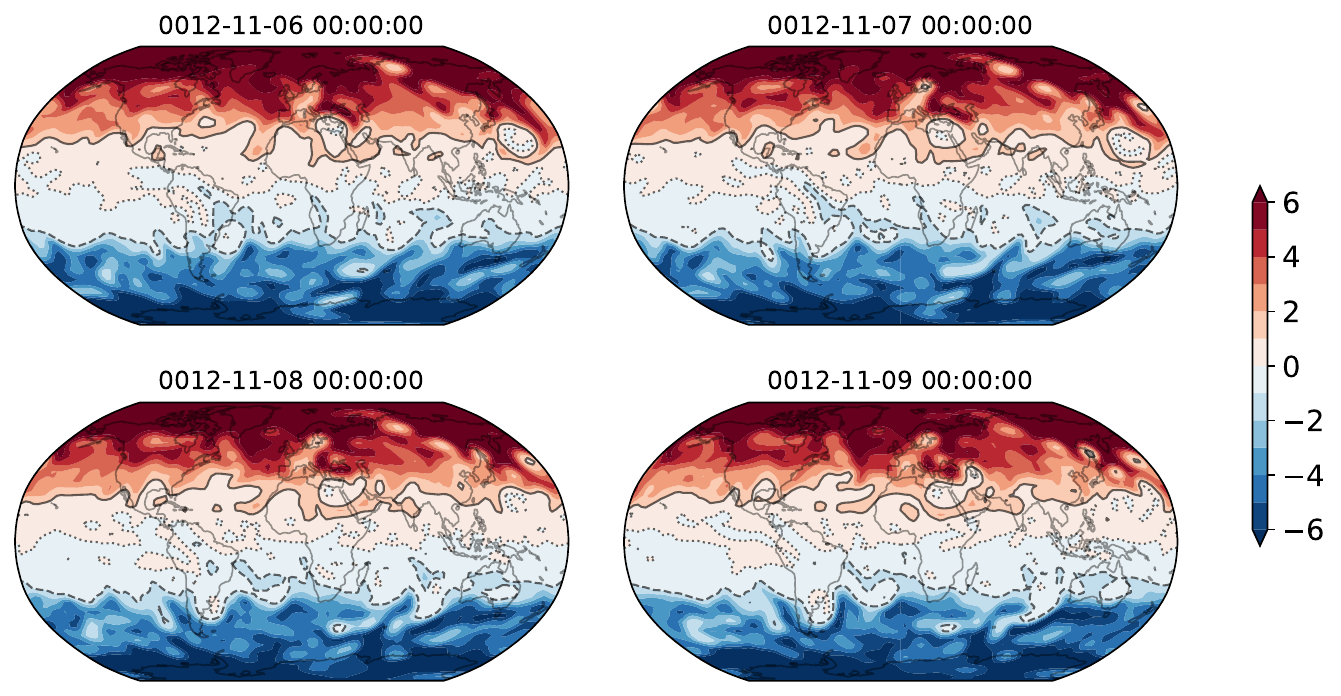}
	\caption{Instantaneous snapshots of Ertel's Potential Vorticity on the 200 mbar surface (colours, in potential vorticity units or PVU) for the $\alpha$=0.0 simulation. Few PVU contours have been marked out separately, for clarity. Continental boundaries are drawn in each panel for reference.}
	\label{fig:flatSST_pv_ut}
\end{figure}

As $\alpha$ decreases, the change in the EMF structure is quite striking. Distinct from the solitary maxima in the simulations with non-zero SST gradient, the EMF has a double maxima structure in the flat SST run, one each in the upper and lower troposphere. The lower tropospheric maximum is weaker and appears to be due to waves generated near the surface that remain confined to the lower troposphere below the 800 mbar level. The upper tropospheric maximum is associated with the waves that are generated near the subtropical tropopause that propagate downwards and equatorwards; these waves cause equatorward eddy heat transport but poleward momentum transport. The Eady growth rate suggests that an upper-level source of baroclinicity is present in all simulations \citep[see Figure S3; ][]{yuval2016eddy}. Waves generated from this upper-level instability become evident when the traditional lower-level baroclinic waves become very weak (note that the scale of the EP flux vectors for the $\alpha$=0.0 run is smallest in Figure \ref{fig:u_ep_t}). A sequence of these waves on the 200 mbar surface is shown in Figure \ref{fig:flatSST_pv_ut}; quite clearly, they ride on the sharp background meridional potential vorticity (PV) gradient and as they evolve, distort and overturn, they act to mix PV across latitudes. A Hovm{\"o}ller diagram of the meridional wind, averaged over 30-40$^{\circ}$N, suggests that these waves have a westward phase speed of $\sim$$2.1$ ms$^{-1}$ and eastward group speed of $\sim$$4.8$ ms$^{-1}$ \citep[not shown; ][]{chang1993}.

\subsection{Zonal mean zonal wind}

The zonal mean zonal wind for the $\alpha$=1.0 simulation (contours in the top left panel of Figure \ref{fig:u_ep_t}) is characterized by a strong subtropical jet, midlatitude westerlies, and low-level tropical easterlies. These features are largely similar to the present-day Earth's atmosphere. However, the upper tropical superrotation\footnote{\textit{Superrotation} refers to zonal mean zonal winds that have an angular momentum greater than the angular momentum of the surface \citep{held_super}.} is distinct from that of the Earth's troposphere. It occurs due to the action of divergent eddy momentum fluxes in the deep tropics \citep[$u'v'|_d$; see Figure S2 ][]{zurita2019role}. In the Earth's upper tropics, deceleration by the seasonally reversing HC exceeds the acceleration by tropical eddies and leads to the prevalent zonal mean easterlies \citep{lee1999climatological, dima2005tropical}. As there is no seasonal cycle, our perpetual equinox simulations superrotate because the mean flow deceleration is too weak to offset the equatorward convergence of momentum by these tropical waves. This leads to the presence of an isolated angular momentum maximum over the deep tropics in all our simulations \citep[Figure \ref{fig:psi_emfd_m};][]{liu_schneider}.

As the imposed SST gradient weakens, lower tropical easterlies and sub-tropical westerlies decrease in strength in concert with the weakening HC intensity. This indicates that there is a decrease in the uptake of angular momentum by the atmosphere from the surface. The thermally-driven subtropical jet and the eddy-driven midlatitude jet begin to separate near $\alpha$=0.4. The subtropical westerly jet disappears altogether for $\alpha$=0.0 run, for which there are distinct columns of zonal mean easterlies in the tropics and westerlies in the midlatitudes, both of which extend from the surface to the tropopause. The deceleration over the equator, and above the surface in the tropics, is achieved by the divergence of rotational eddy momentum flux (see Figures \ref{fig:psi_emfd_m} and S2). In contrast to the other runs, the vertically integrated transport of momentum by the eddies is poleward even in the tropics (Figure S4), confirming the eddy-driven nature of the tropical easterlies in the flat SST run. In accord with \citet{horinouchi}, the EP flux structure for the globally uniform SST case suggests that convergence of momentum by both upper and lower tropospheric eddies is important for the midlatitude westerly flow. Although the imposed SST is globally uniform and the near-surface temperature gradients are very weak, the small vertical shear in the zonal wind profile near the extratropical surface (Figure \ref{fig:u_ep_t}) is sufficient for the near-surface baroclinicity to be important (see the Eady growth rate in Figure S3).

\subsection{Thermal structure}

Colours in Figure \ref{fig:u_ep_t} show the latitudinally anomalous zonal mean temperature. The tropics remain relatively warmer than the other regions for $\alpha$$\ge$0.2, mimicking the imposed SST distribution because the fixed-SST boundary condition strongly constrains the lower tropospheric thermal structure \citep{singhkuang2016}. However, the subtropics become anomalously warm in the flat SST simulation.
The extratropical mean flow and transient waves flux DSE equatorward; this leads to a net accumulation of heat near the HC edge (Figure S1). As the zonal mean DSE transport is dominated by the upper-level circulation \citep{suhas2021}, the warm region is confined to the middle and upper troposphere around the HC edge (see Figure \ref{fig:psi_emfd_m}).

The solid red lines in Figures \ref{fig:psi_emfd_m} and \ref{fig:u_ep_t} show the tropopause height estimated using the WMO definition \citep{reichler2003determining}. In the present-day Earth and in the $\alpha=1.0$ case, while the tropical tropopause height is set by moist convection, the extratropical tropopause is set by the depth of baroclinic eddies \citep{schneider_tropopause}. The tropical tropopause height remains constant as $\alpha$ decreases, but the extratropical tropopause rises; the tropopause is almost flat for $\alpha$=0.0. This can also be seen in the vertical profiles of temperature (Figure S5). The near coincidence of the tropical-extratropical thermal stratifications suggests that convection becomes increasingly important in extratropical dynamics for the runs with flatter SSTs \citep{ait2015eddy,suhas2021}. Further, the lower tropospheric slope of the $\theta_e$ profiles shows a tendency towards increasing convective instability with decreasing SST gradients (Figure S5). 

Our simulations also exhibit a meridionally inverted temperature gradient in the upper troposphere lower stratosphere (UTLS) region, where the tropics are cooler than the poles (Figure \ref{fig:u_ep_t}). Such a temperature distribution arises due to the vertical thermal stratification of the atmosphere. Typically, temperature decreases with height in the troposphere and increases with height in the stratosphere, with a temperature inversion at the tropopause. In all our simulations, the tropical tropopause is slightly higher than in the extratropics. Thus, at $\sim$170 or 180 mbar, where the tropics are still cooling with height, the extratropics have started warming (Figure S5). Hence, the tropics are cooler than the poles in the UTLS region. This ubiquitous UTLS temperature gradient is consistent with the upper-level source of baroclinicity discussed in Section \ref{sec:emf_epf}.

\section{Role of the eddies in driving the overturning circulation}

As described in Section \ref{sec:KE_decomp}, we bring out the role of the eddies by decomposing the eddy-permitting HC into an axisymmetric response and the changes induced by the eddies. The axisymmetric response is captured from another set of simulations where the model setup is the same as the eddy-permitting one, except that the model is truncated at the zonal wavenumber $k$=0. 

\begin{figure}[t!]
	\centering
	\includegraphics[width=0.95\linewidth]{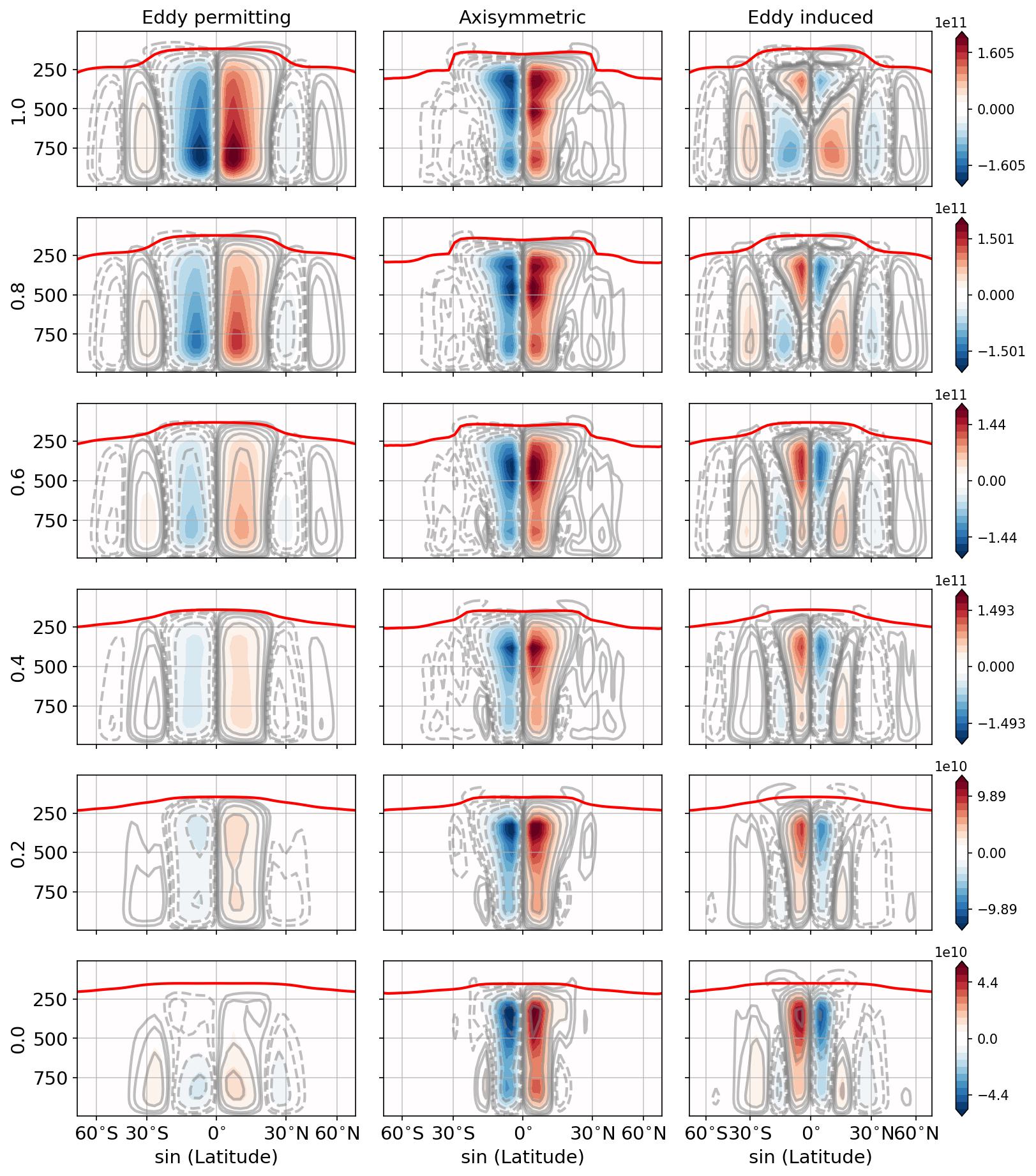}
	\caption{Eddy-permitting (left), axis-symmetric (middle) and eddy-induced (right) streamfunctions estimated from the Kuo-Eliassen equation (Section \ref{sec:KE_decomp}) for different SST profiles. The eddy-induced circulation has been estimated following the methodology of \citet{kim2001hadley}. Contours in light shade are drawn additionally to draw attention to the weaker features of the overturning circulation. The contours are logarithmic in nature, with the lowest magnitude being 10$^{9}$ kg-s$^{-1}$, and the magnitude doubles for successive contours. The red curve in each panel of the first two columns indicates the tropopause height \citep{reichler2003determining} for the respective run. The third column shows the tropopause height for the eddy-permitting run.  
	}
	\label{fig:me_decomp_psi}
\end{figure}

The eddy-permitting, axisymmetric, and eddy-induced cells for varying SST gradients are shown in the first, second, and third columns of Figure \ref{fig:me_decomp_psi}. In the tropics, for $\alpha$=1.0, the axisymmetric circulation is stronger in the upper troposphere, while the eddy-permitting one is broader and more intense in the lower troposphere. Thus, the eddies strengthen the axisymmetric circulation in the lower troposphere but weaken it in the upper troposphere (top row of Figure \ref{fig:me_decomp_psi}). As $\alpha$ decreases, the axisymmetric circulation becomes dominant in the tropics. Further, the eddy-induced circulation has oppositely directed shallow upper tropospheric cells; with decreasing $\alpha$, these cells extend lower down to the surface, resulting in the eddy-driven circulation that opposes the axisymmetric circulation.

\begin{figure}[t!]
	\centering
	\includegraphics[width=\linewidth]{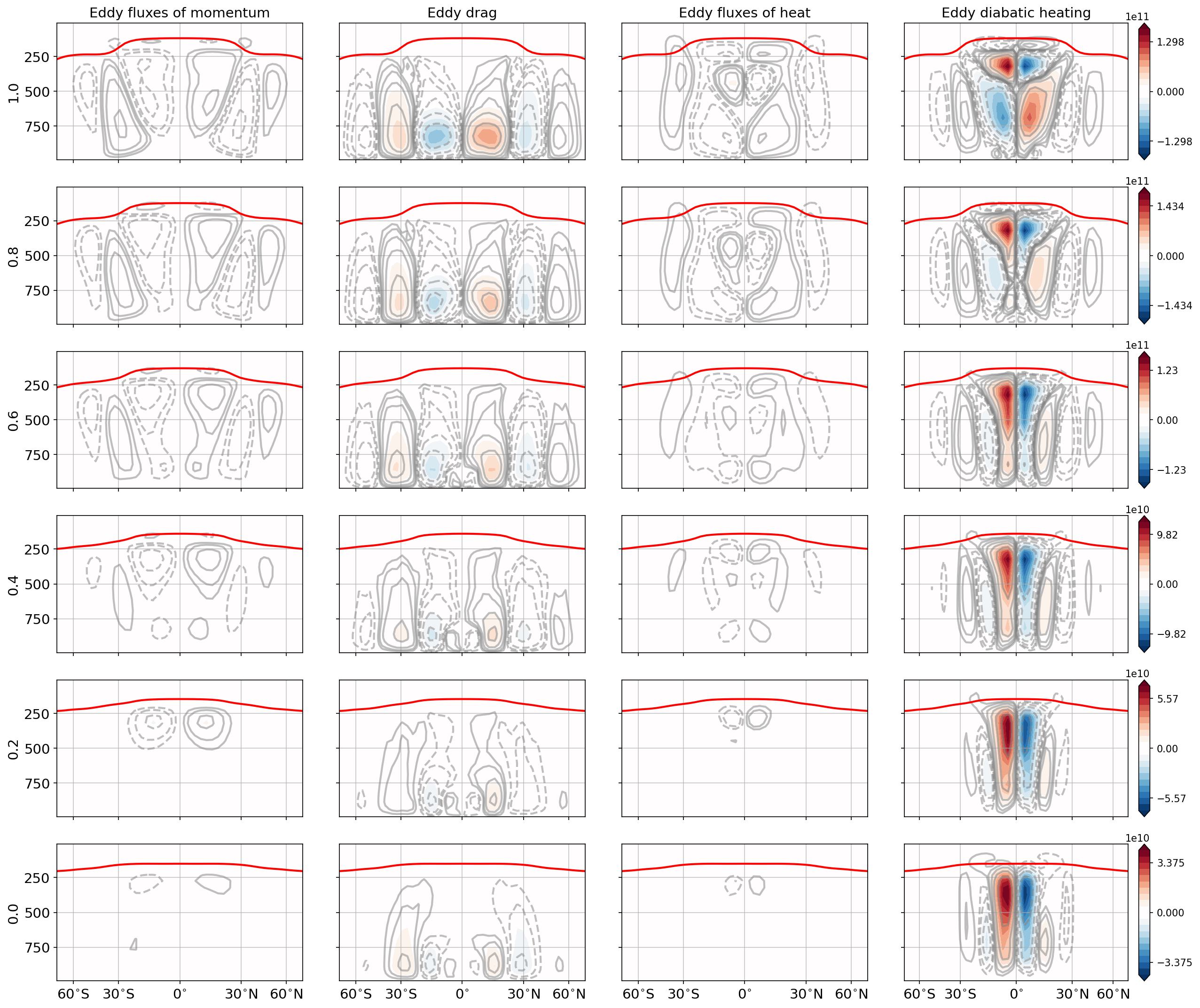}
	\caption{Decomposition of the impact of eddies on the overturning streamfunction for different SST profiles. Contours in light shade are drawn additionally to draw attention to the weaker features of the overturning circulation. The contours are logarithmic in nature, with the lowest magnitude being 10$^{9}$ kg-s$^{-1}$. The magnitude doubles for successive contours. The red curve in each panel indicates the tropopause height  \citep{reichler2003determining} for the eddy-permitting run. 
	}
 \label{fig:psi_ke_decomp}
\end{figure}

\begin{figure}[t]
	\centering
	\includegraphics[width=0.8\linewidth]{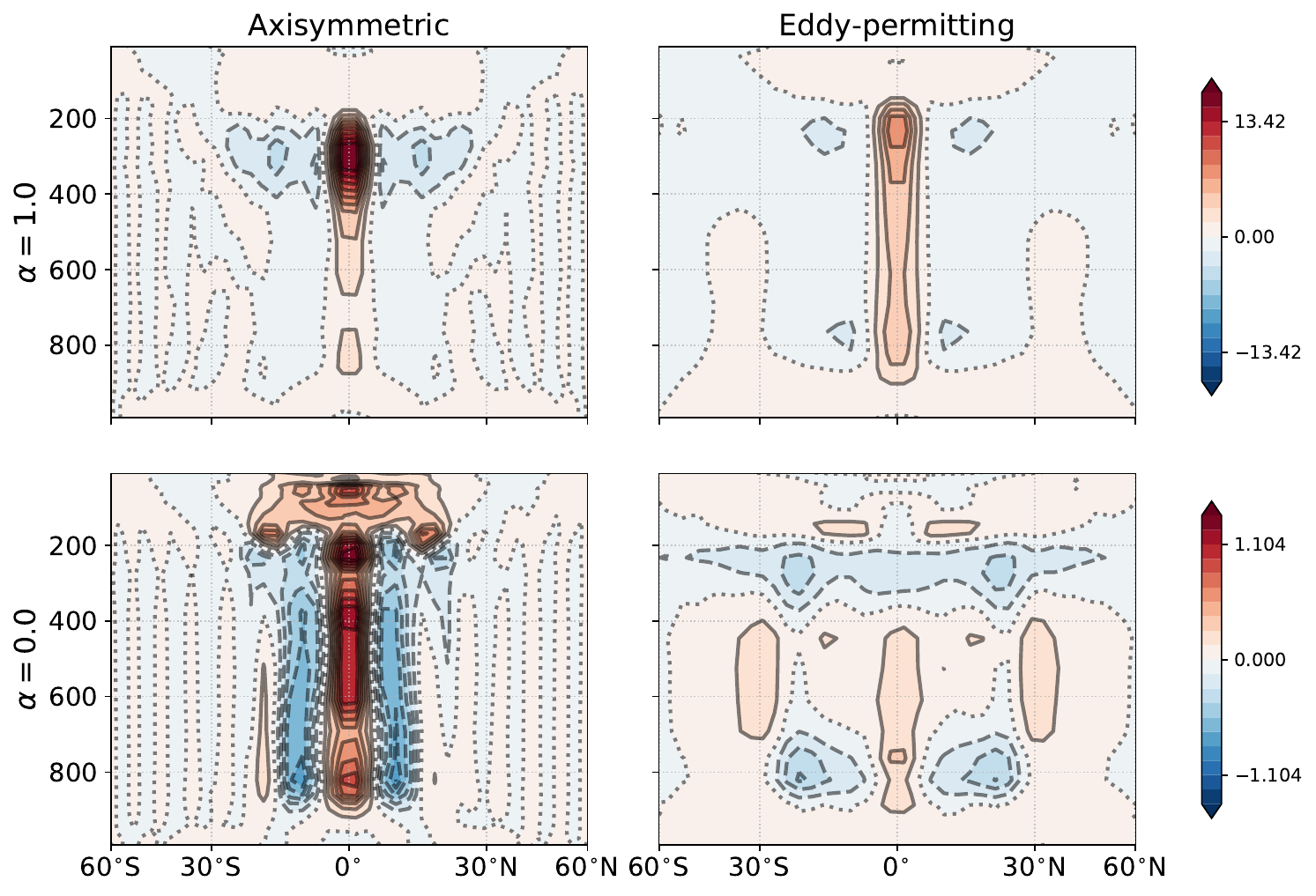}
	\caption{Zonal mean profiles of the diabatic heating rate in K-day$^{-1}$ for (left) axisymmetric and (right) eddy-permitting runs for the (top)$\alpha$=1.0 and (bottom)$\alpha$=0.0 runs. The diabatic heating field is estimated as a residual from the thermodynamic energy budget.
	}
 \label{fig:diabatic}
\end{figure}

To understand the eddy-induced circulation in greater detail, we decompose it into the circulations forced by the eddy heat flux, eddy momentum flux, drag, and diabatic heating terms (see Equation \ref{eq:kim-lee} and Section \ref{sec:KE_decomp}). These are shown in Figure \ref{fig:psi_ke_decomp}, and the diabatic heating rate estimated as a residual from the thermodynamic budget is shown in Figure \ref{fig:diabatic}. Note that the summation of all the columns of Figure \ref{fig:psi_ke_decomp} gives the right-hand column of Figure \ref{fig:me_decomp_psi}. Clearly, the direct influence of eddy heat and momentum fluxes is secondary; the nature of the entire eddy-driven cell in the tropics follows the eddy-induced diabatic heating term. The diabatic heating field (Figure \ref{fig:diabatic}) for the $\alpha$=1.0 axisymmetric run has a large isolated peak in the upper troposphere, while that for the eddy-permitting run is somewhat moderate in intensity in the upper troposphere but extends down to the lower troposphere. Here, eddy heat fluxes modify the diabatic heating field and weaken the meridional temperature gradient, which results in a weaker eddy-permitting HC in the upper troposphere. Further, eddies distribute the diabatic heating over a deeper layer, thereby strengthening the HC in the lower troposphere (Figure \ref{fig:psi_ke_decomp}). Such behavior is consistent with \citet{kim2001hadley}, who showed that the presence of eddies strengthens the HC below, and weakens it above the level of maximum convective heating. This is also consistent with \citet{becker1997feedback}, who showed that the axisymmetric HC is stronger than the eddy-permitting one in the presence of prescribed convective heating in the tropics and weaker in the absence of it. Further, midlatitude eddies flux momentum out of the tropics, decelerating the flow and making it more easterly; this increases the surface drag. As the approximate zonal momentum balance near the surface is between the Coriolis force and the surface drag, a direct cell is induced to offset the effect of enhanced drag (Figure \ref{fig:psi_ke_decomp}).

With weaker imposed SST gradients, the isolated upper tropospheric maxima of diabatic heating in the axisymmetric runs weakens and spreads over the depth of the troposphere (Figure \ref{fig:diabatic}). Thus, the "reverse" cells confined to the upper troposphere for $\alpha$=1.0 systematically descend and displace the direct lower tropospheric circulation (Figures \ref{fig:me_decomp_psi} and \ref{fig:psi_ke_decomp}). The thermally indirect tropical circulation induced by the eddies opposes the direct axisymmetric component; the resultant eddy-permitting circulation is thermally direct and is weaker than the individual axisymmetric and eddy-induced components. We also note the eddy-driven thermally indirect Ferrel cell that is centered near $30^\circ$ and a direct cell further poleward. Indeed, in this Eulerian mean representation, for all $\alpha$, the overturning flow poleward of the subtropics is almost entirely accounted for by the eddy component.

To summarize so far, our axis-symmetric comparisons and the Kuo-Eliassen decomposition clearly show that eddies play an important role in shaping the strength and structure of the HC. However, eddies can be of tropical or extratropical origin and may exert their influences on the HC differently. Next, we will explicitly consider the impact of midlatitude eddies on the HC followed by the impact of tropical eddies.

\subsection{Impact of midlatitude eddies}\label{sec:midlat_eddy}

\begin{figure}[t!]
	\centering
	\includegraphics[width=\linewidth]{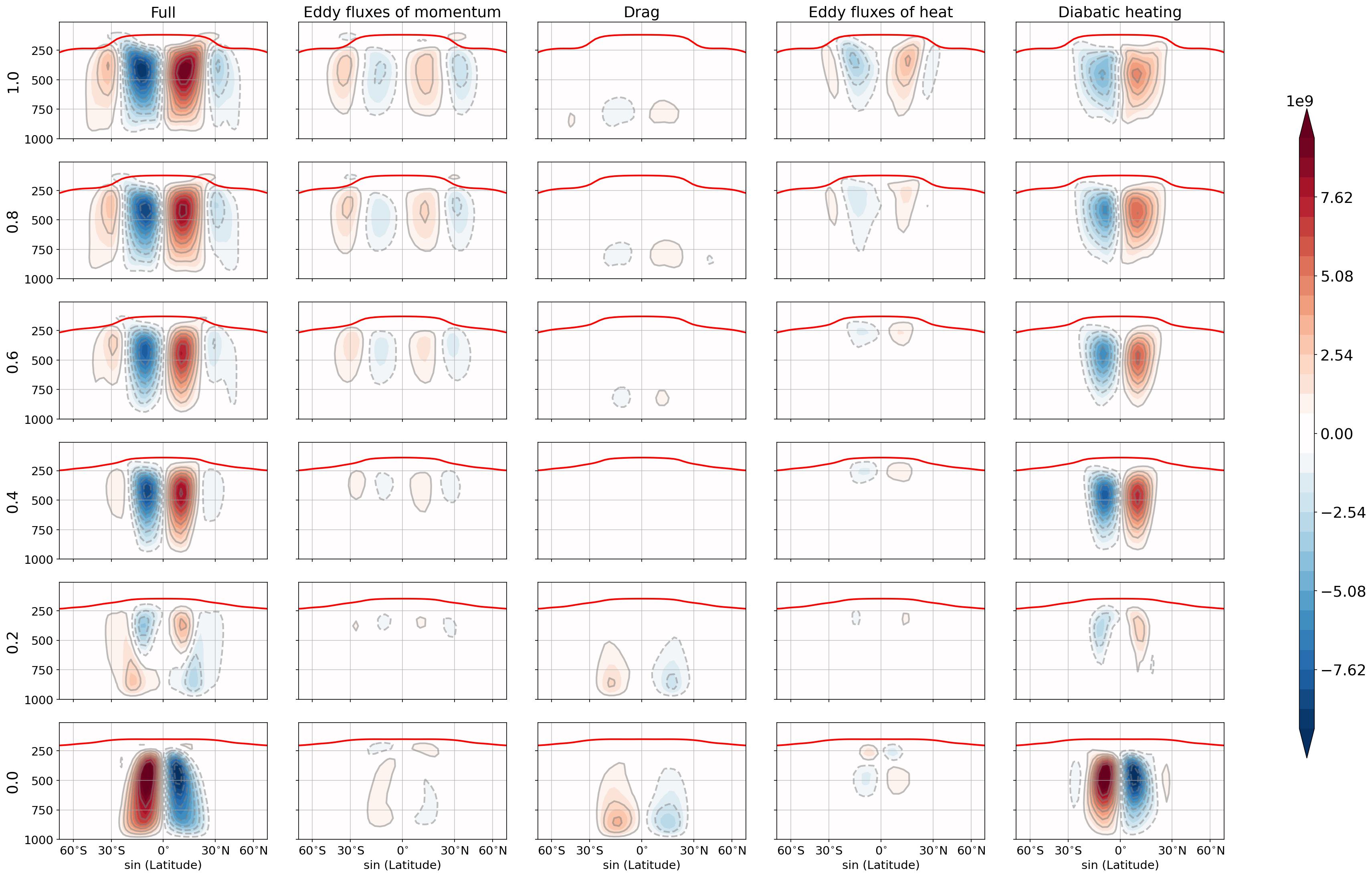}
	\caption{Temporally varying Kuo-Eliassen streamfunction regressed against $u'v'_{max}$ time series. 
	\label{fig:psi_ke_emf_regress}
	}
\end{figure}

As eddy momentum fluxes tend to maximize at the edge of the HC where Rossby waves break \citep{vallis2015}, we use the maximum value of the EMF to construct an index of extratropical eddy variability \citep{zurita2018coupled}. We form a standardized time-series of $\frac{u'v'_{NH,max}-u'v'_{SH,min}}{2}$, and regress the temporally varying streamfunction estimated from the Kuo-Eliassen decomposition (Equation \ref{eq:kuo-eliassen}) against it. This isolates the co-variability between the HC and extratropical baroclinic Rossby waves. The extrema of the EMF are computed over all latitudes in each hemisphere, over all levels, and at every instance of the model output. For reference, the time-averaged position of the NH EMF extrema has been plotted in the rightmost column of Figure S4. Using the extrema from both NH and SH yields a hemispherically symmetric picture of the mid-latitude eddy influence. Note that, as the index of midlatitude variability is uncorrelated with the axisymmetric component, regressing Equation \ref{eq:kim-lee} against the index yields the same results as that for Equation \ref{eq:kuo-eliassen}. 

Figure \ref{fig:psi_ke_emf_regress} shows the results of the regression. As eddy momentum fluxes tend to be largest in the upper troposphere \citep{ait2015eddy}, the near-surface correlation between the regression index and the streamfunction is poor. For conventional SSTs, the midlatitude-eddy-induced circulation (left column of Figure \ref{fig:psi_ke_emf_regress}) acts to strengthen the prevalent time-mean HC (left column of Figure \ref{fig:me_decomp_psi}). This effect gradually weakens with weaker SST gradients, and the midlatitude-eddy-induced circulation abruptly reverses when the imposed SST is globally uniform. These results also suggest that, for non-zero SST gradients, the midlatitude-eddy-induced circulation opposes the full-eddy-induced circulation (right column of Figure \ref{fig:me_decomp_psi}). Further, the midlatitude-eddy-induced circulation is always at least an order of magnitude smaller than both the time-mean HC and the full-eddy-induced circulation; the weak feedback may be due to the absence of stationary eddies in our simulations which can have a large feedback onto the Hadley circulation \citep{becker2001interaction}.

Further examination of the midlatitude-eddy-driven response for the larger $\alpha$ runs, suggests that the regression captures the co-variability between the Hadley and Ferrel cells as well \citep{zurita2018coupled}. As noted above, for conventional SSTs, the midlatitude eddies induce a circulation in the same sense as the prevalent HC (Figure \ref{fig:psi_ke_emf_regress}; first row). The reinforcing nature of this response extends through to all components of Equation \ref{eq:kuo-eliassen}. Midlatitude baroclinic eddies diverge momentum away from the subtropics where they break; this eddy momentum forcing can be balanced by the Coriolis force on the mean meridional circulation, thereby inducing a direct circulation in the tropics \citep[see also][]{held_phillips}. As discussed in Section 4, these eddies modify the heat budget; dynamical cooling in the subtropics due to the poleward fluxing of heat by the eddies is balanced by the adiabatic heating in the sinking branch of the mean flow response, again inducing a direct circulation. 

The strength of the induced circulation weakens with smaller SST gradients. The extratropical response decreases while the tropical response becomes largely dominated by the diabatic heating component. For $\alpha$=0.2, a drag-dominated indirect lower tropospheric response emerges that has a magnitude comparable to that of the upper tropospheric direct circulation. In the runs with weaker SST gradients, eddies that originate in the midlatitudes are able to propagate into the deep tropics (see EP fluxes in Figure \ref{fig:u_ep_t}). They generate a negative momentum forcing right above the equator that is brought down to the surface by this reversed circulation. This negative eddy momentum forcing then aids the uptake of angular momentum from the surface. When the SST becomes globally uniform, the lower tropospheric indirect component strengthens, but the upper tropospheric diabatic component abruptly transitions into a strong oppositely signed response. This response in the globally uniform SST case is coherent with the EP flux vectors that descend downwards from the tropopause, indicating equatorward heat transport by the eddies (Figure \ref{fig:u_ep_t}). The mean flow response offsets the heating perturbation with ascending motion that causes cooling.

Though the regression is aimed at the midlatitude waves, the results in this section indirectly also highlight the importance of the tropical waves for the mean tropical circulation. This is most readily seen in the globally uniform SST case. Since the midlatitude-eddy-induced HC tends to oppose the time-mean HC, the influence of the tropical waves must dominate. Along with the noted differences between the full eddy-induced circulation and the midlatitude-eddy-induced circulation in simulations with non-zero SST gradients, this makes a compelling case for the tropical wave influence on the Hadley circulation. We elucidate the role of the tropical waves in Hadley circulation in the next section.

\subsection{Impact of tropical eddies}

\begin{figure}[t!]
	\centering
	\includegraphics[width=\linewidth]{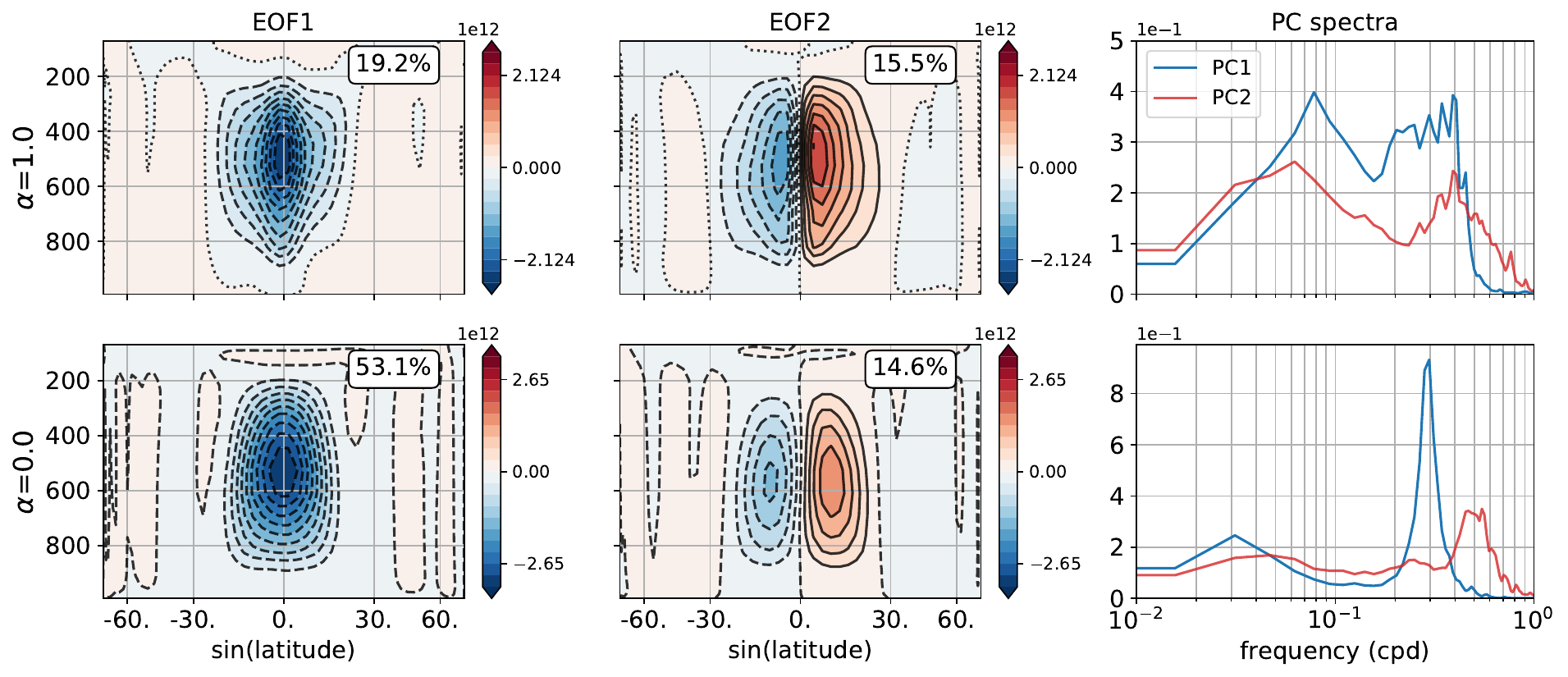}
	\caption{First two empirical orthogonal functions (EOFs) of the time-varying zonal mean overturning streamfunction ($\Psi$, Eq. \ref{eq:mwsf}) for eddy-permitting runs, and variance preserving spectra of the corresponding principal components (PCs) for (top) $\alpha=1.0$ and (bottom) $\alpha=0.0$. The percentage variance explained by the respective EOFs is mentioned in the corresponding panel. Prior to the EOF decomposition, the midlatitude eddy signal was regressed out using the time series from Section \ref{sec:midlat_eddy}, and the streamfunction values were weighted by the square root of the cosine of latitude along the horizontal and by the incremental change in pressure at each pressure level along the vertical \citep{dima2003seasonality}. The PC spectra were computed using Welch's method \citep{welch}. 
	}
	\label{fig:me_eof}
\end{figure}

\begin{figure}[t!]
	\centering
	\includegraphics[width=\linewidth]{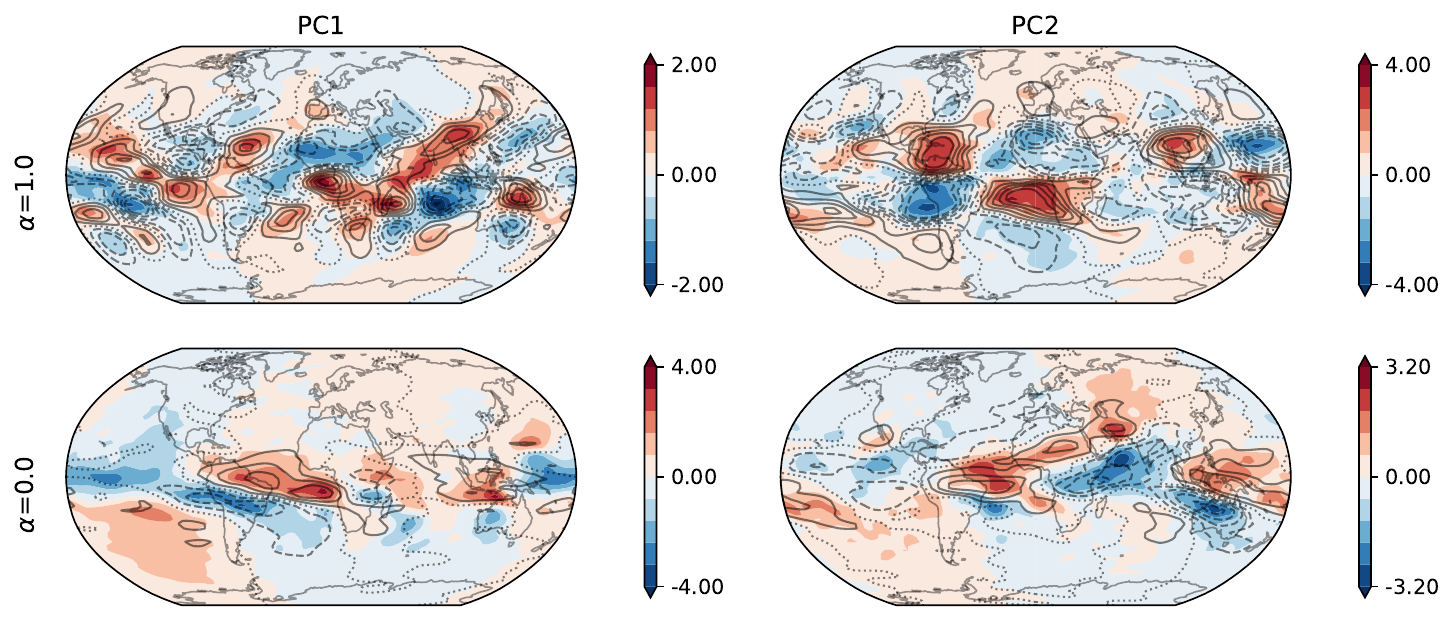}
\caption{Composite maps of the upper-level divergent meridional wind for (top) $\alpha$=1.0 and (bottom) $\alpha$=0.0 simulations. Composites are made over a period where PC1 (left) and PC2 (right) attain a large value. In each panel, colours show the unfiltered anomaly while the contours show the tropical wave-filtered anomalies; solid contours indicate positive values while dashed contours indicate negative values, and the zeroth contour is dotted. In each panel, the contours are such that they contain eleven values, are centered at zero, and span half of the range indicated by the respective colour bar.
 The wave-filtering is performed to isolate the Kelvin wave, mixed Rossby Gravity wave, and the MJO-like mode, consistent with the wavenumber-frequency spectra (Figure S11) and the PC spectra (Figure \ref{fig:me_eof}). The filtering is performed prior to the compositing and over a longer period that includes the compositing period, consistent with \citet{wheeler1999convectively}. Continental boundaries are drawn in each panel for reference.
	}
	\label{fig:me_pc_composites}
\end{figure}

As discussed in the Introduction, convectively coupled equatorial waves can modulate the HC through their influence on tropical precipitation and the associated latent heat release. Given the influence of midlatitude baroclinic waves noted above, the HC must exhibit variability on time scales that are typically associated with equatorial waves along with the synoptic scale variability associated with midlatitude waves. Regression onto an index of tropical variability \citep[wave-filtered precipitation or outgoing longwave radiation as in, for example,][]{WKW} did not yield interpretable results. Hence, the tropical wave influence on the HC is assessed via an Empirical Orthogonal Function (EOF) decomposition of the temporally varying zonal mean overturning streamfunction \citep{dima2003seasonality}. To facilitate a direct attribution to equatorial waves, the midlatitude eddy signal has been regressed out prior to the EOF decomposition.

Figure \ref{fig:me_eof} shows the first two EOFs and the temporal spectra of the associated Principal Components (PCs) for the Earth-like and the globally uniform SST simulations. The most striking feature of the EOFs is that the dominant mode of variability is a hemispherically asymmetric cross-equatorial cell. This mode is somewhat similar to the cross-equatorial variability of the real-world HC associated with the seasonal cycle \citep{dima2003seasonality}. In both simulations, this mode is associated with prominent peaks in both the synoptic and intraseasonal timescales. Specifically, EOF1 for the $\alpha$=1.0 run shows a prominent 30-day peak along with synoptic-scale peaks between 7 and 9 days. The PC for $\alpha$=0.0 has a large peak at the 8-day period along with a secondary 80-day peak. The hemispherically asymmetric mode explains about 20\% of the variance for the Earth-like $\alpha$=1.0 run but accounts for more than 50\% of the variance for the globally uniform SST case. Thus, as posited, the imprint of tropical waves is seen in the timescales of variability of the HC. This also suggests that, even in the Earth-like simulation with equatorially symmetric SSTs and perpetual equinox conditions, the tropical circulation is quite transient, and the equatorial symmetry associated with the Eulerian mean HC (Figure \ref{fig:psi_emfd_m}) is likely a result of temporal averaging. 

In both simulations, the second EOF is equatorially symmetric and explains a similar amount of variance ($\sim$15\%) in both cases. Much like EOF1, the PCs of the second mode also exhibit power in the synoptic and intraseasonal bands, but they are less pronounced. In general, the peaks in the PC spectra can be due to multiple waves present in the equatorial waveguide (see Wheeler-Kiladis spectra in Figure S7). In particular, for $\alpha=0$, the synoptic-scale peaks may be due to the Mixed Rossby Gravity wave or the Kelvin wave, whereas the intraseasonal peaks are due to the MJO-like mode that is ubiquitous in all our simulations (Figure S7).

Apart from their signature in the EOFs, the influence of tropical waves in driving the variability in the tropical overturning circulation is clarified using composite maps of unfiltered and tropical-wave-filtered divergent meridional wind (v$_d$) anomalies shown in Figure \ref{fig:me_pc_composites}; the unfiltered anomalies are shown in colours while the filtered anomalies are shown in contours. We consider the divergent component of the meridional wind because the rotational component would sum to zero when integrated over a latitude circle \citep{zhang2013interannual, sun2019regional}; also, the zonal mean overturning circulation is an ensemble of such localized divergent circulations. The composite is constructed during a period when the respective PCs attained a high value. Figure \ref{fig:me_pc_composites} shows that the unfiltered and tropical wave-filtered v$_d$ anomalies have a close resemblance. 
When the composite is constructed using PC2 for the $\alpha$=1.0 run, the anomalies are characterized by a quadrupole-like structure straddling the equator and are possibly associated with MJO-like convection \citep{monteiro2014interpreting}. This is consistent with the observations of \citet{schwendike}; they showed that the phase-dependent fluctuations of MJO-related convection can modulate the local Hadley and Walker circulations.
For $\alpha$=0.0, the tilted planetary-scale outflows dominate the tropics in the composites made using both PC1 and PC2. Further, for all the SST cases, the pattern correlation between the full anomalies and the tropical-wave-filtered anomalies lies between 0.7 and 0.8.

As the divergent outflows result from convection, v$_d$ anomalies are closely linked with precipitation anomalies. Tropical waves can modulate the precipitation field in the tropics (see the latitudinal variability of the wave-filtered precipitation field in Figure S6) and, thus, influence the ascending branch of the HC through low-level convergence.
The tropical wave influence on the HC, through off-equatorial ascent, is most clearly seen in the hemispherically asymmetric modes. 
Their influence on the symmetric modes (EOF2 of $\alpha$=1.0 \& 0.0) can be visualized clearly when considering these equatorially symmetric EOFs with a flipped sign, as the sign of the EOF is arbitrary. 
Further, the equatorward flux of divergent eddy momentum by the MJO-like mode (see the discussion in Sections 3.2 and 3.3) also triggers a compensatory mean flow response in the same sense as the conventional HC.

\subsection{The Transformed Eulerian Mean Circulation}

\begin{figure}[t]
	\centering
	\includegraphics[width=\linewidth]{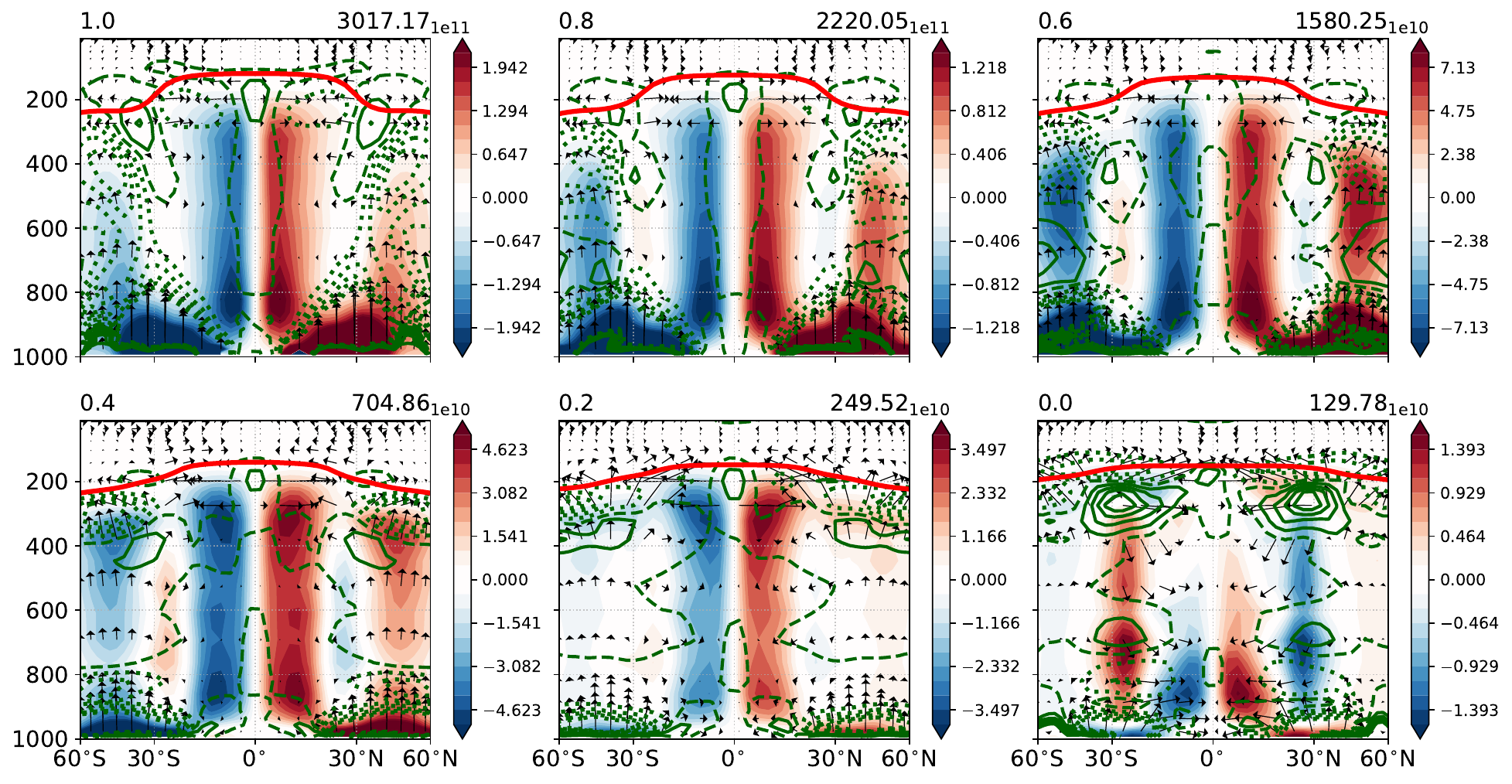}
	\caption{EP flux vectors (quivers), its divergence (contours) and the residual circulation (colours) for the moist eddy-permitting runs. The EP flux divergence is normalised to constrain the range between $-1$ and $1$. 
	}
	\label{fig:tem}
\end{figure}

To tie things together, we now employ the Transformed Eulerian Mean formalism \citep[TEM; ][]{andrews1976} to develop a complementary picture of the zonal mean atmospheric circulation that accounts for the motion induced by eddies. The time mean residual circulation (Figure \ref{fig:tem}) is defined as the conventional mean meridional circulation corrected by the quasi-linear Stokes drift due to atmospheric waves \citep{becker2012}; it closely approximates the diabatic circulation, especially in the upper troposphere and lower stratosphere \citep{held_schneider, juckes, birner2010residual}. For the usual pole-to-equator SST profile, the residual circulation shows a direct hemisphere wide cell as is true of the troposphere \citep{holton2012book}. In the midlatitude lower troposphere, the action of eddies can be seen as the residual circulation abruptly transitions from vertical descent ($\omega^*>0$) to quasi-horizontal polewards flow. Here, the residual circulation and eddy fluxes are strongly coupled as the Coriolis torque on the residual circulation balances the EP flux convergence \citep{birner2010residual}. As discussed in \citet{held_schneider}, the TEM exhibits an infinite return flow near the surface because the eddy heat flux vanishes only in a very thin layer that's unresolved in global climate models. Another reason for such behaviour could be that the quasi-linear Stokes' drift contains the static stability in the denominator, which becomes very small in the boundary layer \citep{held_schneider}.

As $\alpha$ decreases, the eddy heat fluxes become small, and the residual flow looks similar to the conventional Eulerian mean. In the case where $\alpha$=0.0, an equatorward $v^*$ is induced by the isolated centers of EP flux divergence in the subtropical upper troposphere. This is reminiscent of the "downward control" phenomena in the stratosphere \citep{vallis2017atmospheric}, but in a reversed sense. In the real atmosphere, upward propagating wave activity deposits zonal momentum, creating a region of EP flux convergence where Rossby waves break. This induces an overturning circulation that connects the wave-breaking region with the frictional bottom boundary \citep{vallis2017atmospheric}. However, for our $\alpha$=0.0 case, the waves are generated locally due to the upper-level instability that leads to a region of EP flux divergence that induces an equatorward mean flow and consequent thermally indirect cells in the subtropics.

Comparing the Eulerian mean and residual circulations for the $\alpha$=0.0 case (Figures \ref{fig:psi_emfd_m} and \ref{fig:tem} respectively), we see that the reversed midlatitude circulation is not an artifact of averaging; this is quite different from the upper-tropospheric branch of the $\alpha$=1.0 Eulerian mean Ferrel Cell that is absent in the residual circulation. For the $\alpha$=0.0 run, the indirect midlatitude cells are present in the isoentropic mass streamfunction too \citep[not shown; ][]{held_schneider}, indicating that both adiabatic heating through latent heat release and mechanical eddy driving are important for their formation and maintenance. In the tropics, the diabatic circulation is limited only to the mid and lower troposphere. The upper-tropospheric tropical cells are artifacts of the Eulerian mean circulation and are probably eddy-driven. This ties in with the Rossby number-based observation that the HC in the globally uniform SST simulation is eddy-driven (see Section 3.1).

\section{Summary and Discussions}

We have examined the tropical circulation using an aquaplanet GCM forced by fixed zonally symmetric SSTs, spatially uniform solar flux, and fixed orbital parameters. Our primary motivation is to isolate the influence of midlatitude and tropical waves in shaping the Hadley circulation. 
To this end, we vary the SST forcing from a present-day Earth-like warm-equator cold-poles configuration to one which is globally uniform, passively controlling the generation of midlatitude waves. 

A conventional HC-like flow is observed in all our experiments, with large-scale ascent over the equator and descent over the subtropics. The circulations transport dry static energy poleward and latent energy equatorward; the mean overturning circulation largely dictates the transport in the equatorial region. The strength and meridional extent of the HC decrease with weakening SST gradients. When forced by a non-zero SST gradient, organized convection preferentially occurs over latitudes with warmer SSTs and sets up a deep tropical cell. With globally uniform SSTs, the necessary equatorial convergence is provided by tropical waves. A vorticity-based bulk Rossby number suggests that the HC transitions from a thermally and eddy-driven regime to one that is completely eddy-driven. In all simulations, Ferrel cells develop in the midlatitudes that have a sense opposite to that of the tropical Hadley circulation.

All our simulations are super-rotating due to the lack of a seasonal cycle and have colder tropics with warmer poles near the tropopause. In simulations with non-zero SST gradients, the tropospheric zonal wind and temperature distributions are similar to that of the Earth. The zonal wind shows subtropical westerlies, prominent midlatitude eddy-driven jets, and tropical easterlies; the temperature distribution mimics the imposed SST, and eddies control the extratropical tropopause height \citep{held_schneider}. With weakening SST gradients, the subtropical westerlies and the midlatitude eddy-driven jets begin to separate, and the extratropical tropopause rises. When the SSTs are globally uniform, easterlies form through the depth of the tropical troposphere with barotropic westerly jets in the subtropics. In this setup, the subtropics are anomalously warmer than the tropics and poles, and convection determines the extratropical tropopause height.

Our simulations support a full spectrum of tropical waves in the equatorial waveguide. These waves modulate convection and rainfall variability in the tropics. The extratropical waves undergo a significant change with SST gradients. For present-day Earth-like SST profiles, as is expected, Rossby waves generated near the extratropical surface propagate upwards toward the tropopause, which acts like a turning surface and guides them meridionally \citep{ait2015eddy}. They transport heat and angular momentum away from the HC edge, controlling the strength and extent of the HC \citep{walker2006eddy, levine-schneider}. As the SST gradients weaken, the strength of these waves decreases. When the SST is globally uniform, traditional Rossby waves stay confined to the lower troposphere, and we find waves generated near the tropopause due to upper-level baroclinic instability. Like traditional Rossby waves, they transport momentum polewards, but their heat flux is equatorward, consistent with the warmer subtropics in this regime. This structure of the midlatitude waves in the flat SST run causes a double maximum in the eddy momentum flux distribution, one each in the upper and lower troposphere.

The Kuo-Eliassen equation is used to diagnose the impact of atmospheric waves on the HC. In particular, we follow \citet{kim2001hadley} and utilize an additional set of simulations where eddies are suppressed artificially. This allows us to consider the tropical overturning circulation as axisymmetric and eddy-modified components \citep{kim2001hadley}. The presence of eddies in the system changes the thermodynamic and momentum balances and modifies the surface friction and diabatic heating fields. The direct effect of eddy heat and momentum fluxes is found to be secondary. In the present-day Earth-like case, eddies strengthen the axisymmetric component of the HC in the lower troposphere but weaken it in the upper troposphere. With weaker SST gradients, the weakening influence of eddies extends throughout the depth of the troposphere. 

We then consider the influence of midlatitude baroclinic eddies and tropical eddies on the HC separately. The heat and momentum transport by midlatitude baroclinic eddies triggers compensatory transports by the mean flow.
In cases where the imposed pole-to-equator SST gradient is large, the induced flow reinforces the HC. When the SSTs are globally uniform, the eddies induce a thermally indirect circulation that is consistent with the equatorward heat transport by waves generated from the upper-level instability. In contrast, equatorial waves modify the precipitation field in the tropics and modulate the HC ascent branch by their influence on latent heating and low-level convergence. Via an Empirical Orthogonal Function decomposition of the temporally varying mass-weighted streamfunction, we find that the leading modes of variability exhibit considerable hemispheric asymmetry. In fact, in the regime of globally uniform SSTs, a dominant cross-equatorial cell accounts for more than 50\% of the total variance. This suggests that the tropical circulation is quite transient, and the equatorial symmetry associated with the Eulerian mean HC (Figure \ref{fig:psi_emfd_m}) is likely a result of temporal averaging. The hemispherically asymmetric modes are associated with the westward-moving Mixed Rossby-gravity wave, the eastward-moving Kelvin, and an MJO-like mode. Composite maps suggest that the upper-level anomalous flow is closely associated with tropical waves, suggesting that tropical waves play an important role in the transient variability of the HC. Further, the equatorward eddy momentum flux transport by the MJO-like mode can also force a mean flow response in the same sense as the prevalent HC.

Finally, we construct the residual circulation to develop a wave-corrected view of the tropical overturning circulation. For the Earth-like case, the residual circulation is a thermally direct hemisphere-wide cell; it closely resembles the Hadley circulation in the tropics but has a "U-shape" in the midlatitudes, indicating that the residual circulation and the eddy fluxes are strongly coupled. In the flat SST case, the tropical residual circulation is shallow, extending only up to the middle troposphere, reaffirming the eddy-driven nature of the HC. The divergence of EP flux associated with upper-level instability induces subtropical cells in a manner opposite to the "downward control" mechanism in the stratosphere. The flat SST simulation serves as an intermediate case to eddy momentum flux exploration in normal and reversed equator-to-pole temperature gradient experiments that are relevant in the present-day Earth and high obliquity exoplanets, respectively \citep{ait2015eddy, kang2019tropical}. Indeed, a natural extension of the present work would be the inverse SST gradient simulations which are relevant in understanding the atmospheric dynamics of high obliquity exoplanets \citep{lobo}.

A possible way to quantitatively estimate the influence of tropical waves on the strength of the HC would be to simulate an atmosphere without them. However, this appears difficult to achieve because these waves are fundamental modes of the tropical atmosphere; they exist due to the Earth's rotation near the equator \citep{matsuno}. \citet{horinouchi} conducted an experiment where the eddy horizontal winds over the deep tropics were preferentially damped; he observed that the simulated HC was weaker (see their Section 4b). However, this approach would also dampen other waves that can propagate into the deep tropics, thereby negating their interaction with the mean flow. As the zonal mean HC is an ensemble of longitudinally localized divergent circulations \citep{zhang2013interannual, sun2019regional}, damping the eddy horizontal winds would dampen these circulations as well. Further, tropical waves are very sensitive as each wave behaves differently to changes in parameters that control the behavior of convection in the model \citep{frierson2007convectively, peatman, suhas2020, suhas2021}. On reducing the trigger for moist convection, \citet{horinouchi} observed a weakening of the HC along with the disappearance of most of the tropical wave spectra except for some long-period variability. On varying the latent heat of water vapour, thereby changing the degree of moist coupling, \citet{suhas2021} made a similar observation regarding tropical wave variability. Further, it has been observed that large-scale modes of low-frequency tropical variability are intimately tied to the background moisture gradient \citep{sukhatme2013, sukhatme2014}. Changing such parameters impacts the circulation non-trivially, thereby making the attribution to equatorial waves more challenging.

Historically, evaluation of the effect of eddies on the HC has been performed with dry models that use Newtonian relaxation towards a radiative equilibrium temperature profile \citep{held_hou, walker2006eddy, bordoni2010regime}. Such an arrangement allows the occurrence of small-scale dry convection at every latitude \citep{satoh1994} and cannot explain the occurrence of the subtropical dry zones \citep{rodwell, rodwell2001}. However, the real atmosphere is moist, and the release of latent heat pushes the temperature stratification towards moist adiabatic, which is stable to dry convection \citep{satoh1994}. In a moist atmosphere where convection releases latent heat, convection and circulation interactively adjust each other \citep{horinouchi}; a dry atmosphere with Newtonian relaxation precludes such an interactive adjustment. In addition, midlatitude storms exhibit greater strength and larger asymmetry in vertical velocity in the presence of moisture \citep{booth, ogorman}, and their feedback onto the mean flow should be closer to reality in our moist simulations. 

The occurrence of zonal mean Hadley circulations in aquaplanets with spatially uniform SST and insolation (both eddy-permitting and axisymmetric) has been known for long \citep{sumi, kirtman, barsugli, horinouchi, pritchard2016mjo, suhas2021}. However, we present some discussion here for completeness. Typically, convection self-organizes into large clusters in idealized models with homogeneous boundary conditions \citep{muller2022}, in non-rotating models with a rectangular domain \citep{bretherton2005}, and even on a non-rotating sphere \citep{arnold2015global}. With rotation, the atmosphere supports equatorial waves that can concentrate the precipitation in the tropics \citep[Inertia Gravity waves in the axisymmetric setting; see][]{chao2004single}. Convection and circulation interactively adjust each other to form a prominent meridional circulation even in the absence of any imposed thermal gradients \citep{horinouchi}. Thereafter, equatorward moisture flux by the mean flow makes the equatorial region a preferred location for sustained deep convection. 

Noticeable differences exist in the simulated circulations between our $\alpha$$=$$0.0$ run and the $L_v$$=$$1.0$ run of \citet{suhas2021}. Their simulations were performed using the Finite Volume dynamical core of CAM 5.3 of CESM 1.2.2 on an f09 grid ($\sim1^{\circ}$). They obtained a steady and deep HC that extended upto 200 mbar (see their Figure 1b), shallow midlatitude jets extending from the surface only upto 500 mbar, and warmer poles than the equatorial region (their Figure 2b); their vorticity-based local Rossby numbers suggest that the HC is partly eddy- and partly thermally-driven. In contrast, our flat SST HC is eddy-driven and appears transient in the upper troposphere; the midlatitude westerly jets are deep but narrow, and the subtropics are anomalously warmer than the other regions. Apart from the known differences between the two dynamical cores \citep{williamson2008equivalent} and possible divergences occurring between the two due to version differences, the Finite Volume core has an unclosed angular momentum budget \citep{toniazzo2020enforcing}. The spurious sinks of angular momentum tend to decelerate the flow and force secondary circulations with the same sign as the HC, whereby the simulated HC in CAM-FV has a larger mass flux \citep{toniazzo2020enforcing}. These spurious sinks of momentum are possibly responsible for the differences in the zonal wind and temperature distribution (through the thermal wind relation) between our $\alpha$=0.0 run and the $L_v$$=$$1.0$ run of \citet{suhas2021}. Another striking difference is in the simulated transient variability beyond the subtropics; Tropical Cyclone (TC)-like vortices dominate the extratropics in their run, whereas our run shows only some such vortices (not shown). This is likely due to inadequate horizontal resolution in our runs, similar to previous studies that also used a coarse horizontal resolution \citep{sumi, barsugli, kirtman, horinouchi}; compared to previous studies, \citet{shi2014large} obtained the TC-like vortices in their simulations using a finer horizontal grid resolution. Consistent with our results, \citet{suhas2021} also observed transient Rossby wave activity in the tropics and subtropics connected with the upper-level instability (personal communication).

The changing nature of the driving of the HC with the weakening SST gradients is of potential interest in greenhouse gas warming scenarios where the decrease in meridional surface temperature gradient is a robust trend \citep{adam, vallis2015}. In addition, the tropical upper troposphere becomes anomalously warm due to a decreasing saturated lapse rate \citep{vallis2015}. This causes the baroclinicity to weaken in the lower troposphere but strengthen in the upper troposphere \citep{wu_baroclinicity, yuval2016eddy}. As the growth rate of baroclinic instability depends on the imposed meridional temperature gradient \citep{eady}, the corresponding baroclinicity and strength of the baroclinic waves in our simulations decrease in concert with weakening gradients of SST \citep{kodama}. 
The upper-level baroclinicity is tied to the near tropopause reverse temperature gradient that's ubiquitous in all our simulations. It is possible that these upper-level waves are present in all our simulations but are overshadowed by the stronger waves generated in the lower troposphere for present-day Earth-like SST profiles. In fact, \citet{horinouchi} obtained a similar upper-level instability but attributed it to the shear instability of the angular momentum conserving zonal wind. In the present-day Earth's atmosphere, prominent regions of EP flux divergence have been observed on the poleward flanks of the jet core in both hemispheres \citep{birner2013up}, and a tropopause-level planetary wave source can trigger waves that force Sudden Stratospheric Warmings \citep{boljka}.

As discussed in the Introduction, an understanding of the zonal mean overturning circulation has developed in two well-separated regimes. While the eddy-mean flow interaction related to baroclinic midlatitude waves has been studied with Earth-like warm equator-cold poles thermal forcing \citep{schneider_review, walker2006eddy, korty-schneider, levine-schneider}, the role of tropical waves has been indirectly elucidated with spatially uniform thermal forcing \citep{sumi, kirtman, horinouchi, barsugli, suhas2021}. Our fixed SST simulations, with the intermediate thermal gradients, connect the two regimes and cover the gap that has existed so far in the literature. Indeed, the atmosphere-ocean is a coupled system, and the atmospheric circulation should be able to modify the SST distribution \citep{xie}. In this regard, fixed SST aquaplanet simulations are an idealization but form an important first step in developing a more complete understanding of the drivers of the tropical overturning circulation \citep{held2005gap, jeevanjee2017, maher2019model}. 

\bibliographystyle{apalike}
\bibliography{ref.bib}

\begin{thebibliography}{}

\bibitem[Adam et~al., 2014]{adam}
Adam, O., Schneider, T., and Harnik, N. (2014).
\newblock Role of changes in mean temperatures versus temperature gradients in
  the recent widening of the hadley circulation.
\newblock {\em Journal of Climate}, 27(19):7450--7461.

\bibitem[Adames and Wallace, 2014]{aw1}
Adames, A. and Wallace, J. (2014).
\newblock Three-dimensional structure and evolution of the mjo and its relation
  to the mean flow.
\newblock {\em Journal of the Atmospheric Sciences}, 71(3):2008--2026.

\bibitem[Ait-Chaalal and Schneider, 2015]{ait2015eddy}
Ait-Chaalal, F. and Schneider, T. (2015).
\newblock Why eddy momentum fluxes are concentrated in the upper troposphere.
\newblock {\em Journal of the Atmospheric Sciences}, 72(4):1585--1604.

\bibitem[Andrews and Mcintyre, 1976]{andrews1976}
Andrews, D. and Mcintyre, M.~E. (1976).
\newblock Planetary waves in horizontal and vertical shear: The generalized
  eliassen-palm relation and the mean zonal acceleration.
\newblock {\em Journal of Atmospheric Sciences}, 33(11):2031--2048.

\bibitem[Arnold and Randall, 2015]{arnold2015global}
Arnold, N. and Randall, D. (2015).
\newblock Global-scale convective aggregation: Implications for the
  madden-julian oscillation.
\newblock {\em Journal of Advances in Modeling Earth Systems}, 7(4):1499--1518.

\bibitem[Barsugli et~al., 2005]{barsugli}
Barsugli, J., Shin, S., and Sardeshmukh, P. (2005).
\newblock Tropical climate regimes and global climate sensitivity in a simple
  setting.
\newblock {\em Journal of the atmospheric sciences}, 62(4):1226--1240.

\bibitem[Becker, 2012]{becker2012}
Becker, E. (2012).
\newblock Dynamical control of the middle atmosphere.
\newblock {\em Space Science Reviews}, 168(1):283--314.

\bibitem[Becker and Schmitz, 2001]{becker2001interaction}
Becker, E. and Schmitz, G. (2001).
\newblock Interaction between extratropical stationary waves and the zonal mean
  circulation.
\newblock {\em Journal of the atmospheric sciences}, 58(5):462--480.

\bibitem[Becker et~al., 1997]{becker1997feedback}
Becker, E., Schmitz, G., and Gepr{\"a}gs, R. (1997).
\newblock The feedback of midlatitude waves onto the hadley cell in a simple
  general circulation model.
\newblock {\em Tellus A}, 49(2):182--199.

\bibitem[Birner, 2010]{birner2010residual}
Birner, T. (2010).
\newblock Residual circulation and tropopause structure.
\newblock {\em Journal of the Atmospheric Sciences}, 67(8):2582--2600.

\bibitem[Birner et~al., 2013]{birner2013up}
Birner, T., J.~Thompson, D., and Shepherd, T. (2013).
\newblock Up-gradient eddy fluxes of potential vorticity near the subtropical
  jet.
\newblock {\em Geophysical research letters}, 40(22):5988--5993.

\bibitem[Boljka and Birner, 2020]{boljka}
Boljka, L. and Birner, T. (2020).
\newblock Tropopause-level planetary wave source and its role in two-way
  troposphere--stratosphere coupling.
\newblock {\em Weather and Climate Dynamics}, 1(2):555--575.

\bibitem[Booth et~al., 2015]{booth}
Booth, J.~F., Polvani, L., O'Gorman, P.~A., and Wang, S. (2015).
\newblock Effective stability in a moist baroclinic wave.
\newblock {\em Atmospheric Science Letters}, 16(1):56--62.

\bibitem[Bordoni and Schneider, 2010]{bordoni2010regime}
Bordoni, S. and Schneider, T. (2010).
\newblock Regime transitions of steady and time-dependent hadley circulations:
  Comparison of axisymmetric and eddy-permitting simulations.
\newblock {\em Journal of the Atmospheric Sciences}, 67(5):1643--1654.

\bibitem[Bretherton et~al., 2005]{bretherton2005}
Bretherton, C.~S., Blossey, P.~N., and Khairoutdinov, M. (2005).
\newblock An energy-balance analysis of deep convective self-aggregation above
  uniform sst.
\newblock {\em Journal of the atmospheric sciences}, 62(12):4273--4292.

\bibitem[Chang, 1993]{chang1993}
Chang, E.~K. (1993).
\newblock Downstream development of baroclinic waves as inferred from
  regression analysis.
\newblock {\em Journal of Atmospheric Sciences}, 50(13):2038--2053.

\bibitem[Chang, 1996]{chang1996}
Chang, E.~K. (1996).
\newblock Mean meridional circulation driven by eddy forcings of different
  timescales.
\newblock {\em Journal of Atmospheric Sciences}, 53(1):113--125.

\bibitem[Chao and Chen, 2004]{chao2004single}
Chao, W. and Chen, B. (2004).
\newblock Single and double itcz in an aqua-planet model with constant sea
  surface temperature and solar angle.
\newblock {\em Climate Dynamics}, 22(4):447--459.

\bibitem[Chemke and Polvani, 2019]{chemke2019opposite}
Chemke, R. and Polvani, L.~M. (2019).
\newblock Opposite tropical circulation trends in climate models and in
  reanalyses.
\newblock {\em Nature Geoscience}, 12(7):528--532.

\bibitem[Danabasoglu et~al., 2020]{danabasoglu2020community}
Danabasoglu, G., Lamarque, J.-F., Bacmeister, J., Bailey, D., DuVivier, A.,
  Edwards, J., Emmons, L., Fasullo, J., Garcia, R., Gettelman, A., et~al.
  (2020).
\newblock The community earth system model version 2 (cesm2).
\newblock {\em Journal of Advances in Modeling Earth Systems},
  12(2):e2019MS001916.

\bibitem[Das et~al., 2016]{das2016low}
Das, S. et~al. (2016).
\newblock Low-frequency intraseasonal variability in a zonally symmetric
  aquaplanet model.
\newblock {\em Meteorology and Atmospheric Physics}, 128(6):697--713.

\bibitem[Davis and Birner, 2019]{davis2019}
Davis, N.~A. and Birner, T. (2019).
\newblock Eddy influences on the hadley circulation.
\newblock {\em Journal of Advances in Modeling Earth Systems},
  11(6):1563--1581.

\bibitem[Deser and Phillips, 2009]{deser}
Deser, C. and Phillips, A.~S. (2009).
\newblock Atmospheric circulation trends, 1950--2000: The relative roles of sea
  surface temperature forcing and direct atmospheric radiative forcing.
\newblock {\em Journal of Climate}, 22(2):396--413.

\bibitem[Dima and Wallace, 2003]{dima2003seasonality}
Dima, I. and Wallace, J. (2003).
\newblock On the seasonality of the hadley cell.
\newblock {\em Journal of the atmospheric sciences}, 60(12):1522--1527.

\bibitem[Dima et~al., 2005]{dima2005tropical}
Dima, I., Wallace, J., and Kraucunas, I. (2005).
\newblock Tropical zonal momentum balance in the ncep reanalyses.
\newblock {\em Journal of the atmospheric sciences}, 62(7):2499--2513.

\bibitem[Eady, 1949]{eady}
Eady, E.~T. (1949).
\newblock Long waves and cyclone waves.
\newblock {\em Tellus}, 1(3):33--52.

\bibitem[Edmon et~al., 1980]{edmon1980eliassen}
Edmon, H., Hoskins, B., and McIntyre, M. (1980).
\newblock Eliassen-palm cross sections for the troposphere.
\newblock {\em Journal of Atmospheric Sciences}, 37(12):2600--2616.

\bibitem[Fedorov et~al., 2010]{fedorov2010tropical}
Fedorov, A., Brierley, C., and Emanuel, K. (2010).
\newblock Tropical cyclones and permanent el ni{\~n}o in the early pliocene
  epoch.
\newblock {\em Nature}, 463(7284):1066--1070.

\bibitem[Fedorov et~al., 2019]{fedorov2019tropical}
Fedorov, A. et~al. (2019).
\newblock Tropical cyclogenesis in warm climates simulated by a cloud-system
  resolving model.
\newblock {\em Climate Dynamics}, 52(1-2):107--127.

\bibitem[Frierson, 2007]{frierson2007convectively}
Frierson, D. (2007).
\newblock Convectively coupled kelvin waves in an idealized moist general
  circulation model.
\newblock {\em Journal of the Atmospheric Sciences}, 64(6):2076--2090.

\bibitem[Geen et~al., 2020]{geen}
Geen, R., Bordoni, S., Battisti, D.~S., and Hui, K. (2020).
\newblock Monsoons, itczs, and the concept of the global monsoon.
\newblock {\em Reviews of Geophysics}, 58(4):e2020RG000700.

\bibitem[Goldblatt and Watson, 2012]{goldblatt}
Goldblatt, C. and Watson, A.~J. (2012).
\newblock The runaway greenhouse: implications for future climate change,
  geoengineering and planetary atmospheres.
\newblock {\em Philosophical Transactions of the Royal Society A: Mathematical,
  Physical and Engineering Sciences}, 370(1974):4197--4216.

\bibitem[Held, 1999]{held_super}
Held, I.~M. (1999).
\newblock Equatorial superrotation in earth-like atmospheric models.
\newblock {\em Bernard Haurwitz Memorial Lecture}.

\bibitem[Held, 2005]{held2005gap}
Held, I.~M. (2005).
\newblock The gap between simulation and understanding in climate modeling.
\newblock {\em Bulletin of the American Meteorological Society},
  86(11):1609--1614.

\bibitem[Held, 2019]{held_review}
Held, I.~M. (2019).
\newblock 100 years of progress in understanding the general circulation of the
  atmosphere.
\newblock {\em Meteorological Monographs}, 59:6--1.

\bibitem[Held and Hou, 1980]{held_hou}
Held, I.~M. and Hou, A.~Y. (1980).
\newblock Nonlinear axially symmetric circulations in a nearly inviscid
  atmosphere.
\newblock {\em Journal of the Atmospheric Sciences}, 37(3):515--533.

\bibitem[Held and Phillips, 1990]{held_phillips}
Held, I.~M. and Phillips, P.~J. (1990).
\newblock A barotropic model of the interaction between the hadley cell and a
  rossby wave.
\newblock {\em Journal of Atmospheric Sciences}, 47(7):856--869.

\bibitem[Held and Schneider, 1999]{held_schneider}
Held, I.~M. and Schneider, T. (1999).
\newblock The surface branch of the zonally averaged mass transport circulation
  in the troposphere.
\newblock {\em Journal of the atmospheric sciences}, 56(11):1688--1697.

\bibitem[Holton and Hakim, 2012]{holton2012book}
Holton, J. and Hakim, G. (2012).
\newblock {\em An Introduction to Dynamic Meteorology}.
\newblock Academic Press, New York, 5 edition.

\bibitem[Horinouchi, 2012]{horinouchi}
Horinouchi, T. (2012).
\newblock Moist hadley circulation: Possible role of wave--convection coupling
  in aquaplanet experiments.
\newblock {\em Journal of the atmospheric sciences}, 69(3):891--907.

\bibitem[Hoskins and Yang, 2021]{hoskins2021}
Hoskins, B. and Yang, G.-Y. (2021).
\newblock The detailed dynamics of the hadley cell. part ii:
  December--february.
\newblock {\em Journal of Climate}, 34(2):805--823.

\bibitem[Hoskins and Yang, 2023]{hoskins2023}
Hoskins, B. and Yang, G.-Y. (2023).
\newblock A global perspective on the upper branch of the hadley cell.
\newblock {\em Journal of Climate}, pages 1--40.

\bibitem[Hoskins et~al., 2020]{hoskins2020}
Hoskins, B., Yang, G.-Y., and Fonseca, R. (2020).
\newblock The detailed dynamics of the june--august hadley cell.
\newblock {\em Quarterly Journal of the Royal Meteorological Society},
  146(727):557--575.

\bibitem[Ingersoll, 1969]{ingersoll}
Ingersoll, A.~P. (1969).
\newblock The runaway greenhouse: A history of water on venus.
\newblock {\em Journal of Atmospheric Sciences}, 26(6):1191--1198.

\bibitem[Jeevanjee et~al., 2017]{jeevanjee2017}
Jeevanjee, N., Hassanzadeh, P., Hill, S., and Sheshadri, A. (2017).
\newblock A perspective on climate model hierarchies.
\newblock {\em Journal of Advances in Modeling Earth Systems}, 9(4):1760--1771.

\bibitem[Juckes, 2001]{juckes}
Juckes, M. (2001).
\newblock A generalization of the transformed eulerian-mean meridional
  circulation.
\newblock {\em Quarterly Journal of the Royal Meteorological Society},
  127(571):147--160.

\bibitem[Kang et~al., 2019]{kang2019tropical}
Kang, W., Cai, M., and Tziperman, E. (2019).
\newblock Tropical and extratropical general circulation with a meridional
  reversed temperature gradient as expected in a high obliquity planet.
\newblock {\em Icarus}, 330:142--154.

\bibitem[Kiladis et~al., 2009]{Kiladis-rev}
Kiladis, G. et~al. (2009).
\newblock Convectively coupled equatorial waves.
\newblock {\em Reviews of Geophysics}, 47(2).

\bibitem[Kim and Lee, 2001]{kim2001hadley}
Kim, H.-k. and Lee, S. (2001).
\newblock Hadley cell dynamics in a primitive equation model. part ii:
  Nonaxisymmetric flow.
\newblock {\em Journal of the Atmospheric Sciences}, 58(19):2859--2871.

\bibitem[Kirtman and Schneider, 2000]{kirtman}
Kirtman, B. and Schneider, E. (2000).
\newblock A spontaneously generated tropical atmospheric general circulation.
\newblock {\em Journal of the atmospheric sciences}, 57(13):2080--2093.

\bibitem[Kodama and Iwasaki, 2009]{kodama}
Kodama, C. and Iwasaki, T. (2009).
\newblock Influence of the sst rise on baroclinic instability wave activity
  under an aquaplanet condition.
\newblock {\em Journal of the atmospheric sciences}, 66(8):2272--2287.

\bibitem[Korty and Emanuel, 2007]{kortyemanuel}
Korty, R. and Emanuel, K. (2007).
\newblock The dynamic response of the winter stratosphere to an equable climate
  surface temperature gradient.
\newblock {\em Journal of Climate}, 20:4213--5228.

\bibitem[Korty and Schneider, 2008]{korty-schneider}
Korty, R.~L. and Schneider, T. (2008).
\newblock Extent of hadley circulations in dry atmospheres.
\newblock {\em Geophysical Research Letters}, 35(23).

\bibitem[Kuo, 1956]{kuo1956forced}
Kuo, H. (1956).
\newblock Forced and free meridional circulations in the atmosphere.
\newblock {\em Journal of the Atmospheric Sciences}, 13(6):561--568.

\bibitem[Lee, 1999]{lee1999climatological}
Lee, S. (1999).
\newblock Why are the climatological zonal winds easterly in the
  equatorialupper troposphere?
\newblock {\em Journal of the atmospheric sciences}, 56(10):1353--1363.

\bibitem[Levine and Schneider, 2015]{levine-schneider}
Levine, X. and Schneider, T. (2015).
\newblock Baroclinic eddies and the extent of the hadley circulation: An
  idealized gcm study.
\newblock {\em Journal of the Atmospheric Sciences}, 72:2744--2761.

\bibitem[Liu and Schneider, 2011]{liu_schneider}
Liu, J. and Schneider, T. (2011).
\newblock Convective generation of equatorial superrotation in planetary
  atmospheres.
\newblock {\em Journal of the atmospheric sciences}, 68(11):2742--2756.

\bibitem[Lobo and Bordoni, 2020]{lobo}
Lobo, A.~H. and Bordoni, S. (2020).
\newblock Atmospheric dynamics in high obliquity planets.
\newblock {\em Icarus}, 340:113592.

\bibitem[Maher et~al., 2019]{maher2019model}
Maher, P. et~al. (2019).
\newblock Model hierarchies for understanding atmospheric circulation.
\newblock {\em Reviews of Geophysics}, 57(2):250--280.

\bibitem[Matsuno, 1966]{matsuno}
Matsuno, T. (1966).
\newblock Quasi-geostrophic motions in the equatorial area.
\newblock {\em Journal of the Meteorological Society of Japan. Ser. II},
  44(1):25--43.

\bibitem[Medeiros et~al., 2016]{medeiros2016reference}
Medeiros, B., Williamson, D., and Olson, J. (2016).
\newblock Reference aquaplanet climate in the community atmosphere model,
  version 5.
\newblock {\em Journal of Advances in Modeling Earth Systems}, 8(1):406--424.

\bibitem[Monteiro et~al., 2014]{monteiro2014interpreting}
Monteiro, J.~M., Adames, {\'A}.~F., Wallace, J.~M., and Sukhatme, J.~S. (2014).
\newblock Interpreting the upper level structure of the madden-julian
  oscillation.
\newblock {\em Geophysical Research Letters}, 41(24):9158--9165.

\bibitem[Muller et~al., 2022]{muller2022}
Muller, C., Yang, D., Craig, G., Cronin, T., Fildier, B., Haerter, J.~O.,
  Hohenegger, C., Mapes, B., Randall, D., Shamekh, S., et~al. (2022).
\newblock Spontaneous aggregation of convective storms.
\newblock {\em Annual Review of Fluid Mechanics}, 54:133--157.

\bibitem[Neale and Hoskins, 2000]{neale}
Neale, R.~B. and Hoskins, B.~J. (2000).
\newblock A standard test for agcms including their physical parametrizations:
  I: The proposal.
\newblock {\em Atmospheric Science Letters}, 1(2):101--107.

\bibitem[O’Gorman, 2011]{ogorman}
O’Gorman, P.~A. (2011).
\newblock The effective static stability experienced by eddies in a moist
  atmosphere.
\newblock {\em Journal of the Atmospheric Sciences}, 68(1):75--90.

\bibitem[Peatman et~al., 2018]{peatman}
Peatman, S.~C., Methven, J., and Woolnough, S.~J. (2018).
\newblock Isolating the effects of moisture entrainment on convectively coupled
  equatorial waves in an aquaplanet gcm.
\newblock {\em Journal of the Atmospheric Sciences}, 75(9):3139--3157.

\bibitem[Peixoto and Oort, 1992]{peixoto1992physics}
Peixoto, J.~P. and Oort, A.~H. (1992).
\newblock {\em Physics of climate}, volume 520.
\newblock Springer.

\bibitem[Pierrehumbert and Swanson, 1995]{pierrehumbert_baroclinic}
Pierrehumbert, R. and Swanson, K. (1995).
\newblock Baroclinic instability.
\newblock {\em Annual review of fluid mechanics}, 27(1):419--467.

\bibitem[Pierrehumbert, 2005]{pierrehumbert_snowball}
Pierrehumbert, R.~T. (2005).
\newblock Climate dynamics of a hard snowball earth.
\newblock {\em Journal of Geophysical Research: Atmospheres}, 110(D1).

\bibitem[Polvani et~al., 2011]{polvani_ozone}
Polvani, L.~M., Waugh, D.~W., Correa, G.~J., and Son, S.-W. (2011).
\newblock Stratospheric ozone depletion: The main driver of twentieth-century
  atmospheric circulation changes in the southern hemisphere.
\newblock {\em Journal of Climate}, 24(3):795--812.

\bibitem[Pritchard and Yang, 2016]{pritchard2016mjo}
Pritchard, M.~S. and Yang, D. (2016).
\newblock Response of the superparameterized madden--julian oscillation to
  extreme climate and basic-state variation challenges a moisture mode view.
\newblock {\em Journal of Climate}, 29(13):4995--5008.

\bibitem[Reichler et~al., 2003]{reichler2003determining}
Reichler, T., Dameris, M., and Sausen, R. (2003).
\newblock Determining the tropopause height from gridded data.
\newblock {\em Geophysical research letters}, 30(20).

\bibitem[Rodwell and Hoskins, 1996]{rodwell}
Rodwell, M.~J. and Hoskins, B.~J. (1996).
\newblock Monsoons and the dynamics of deserts.
\newblock {\em Quarterly Journal of the Royal Meteorological Society},
  122(534):1385--1404.

\bibitem[Rodwell and Hoskins, 2001]{rodwell2001}
Rodwell, M.~J. and Hoskins, B.~J. (2001).
\newblock Subtropical anticyclones and summer monsoons.
\newblock {\em Journal of Climate}, 14(15):3192--3211.

\bibitem[Satoh, 1994]{satoh1994}
Satoh, M. (1994).
\newblock Hadley circulations in radiative--convective equilibrium in an
  axially symmetric atmosphere.
\newblock {\em Journal of Atmospheric Sciences}, 51(13):1947--1968.

\bibitem[Satoh et~al., 1995]{satoh1995}
Satoh, M., Shiobara, M., and Takahashi, M. (1995).
\newblock Hadley circulations and their r{\^o}les in the global angular
  momentum budget in two-and three-dimensional models.
\newblock {\em Tellus A}, 47(5):548--560.

\bibitem[Schneider, 1977]{schneider1977}
Schneider, E.~K. (1977).
\newblock Axially symmetric steady-state models of the basic state for
  instability and climate studies. part ii. nonlinear calculations.
\newblock {\em Journal of Atmospheric Sciences}, 34(2):280--296.

\bibitem[Schneider and Lindzen, 1977]{schneider_lindzen}
Schneider, E.~K. and Lindzen, R.~S. (1977).
\newblock Axially symmetric steady-state models of the basic state for
  instability and climate studies. part i. linearized calculations.
\newblock {\em Journal of Atmospheric Sciences}, 34(2):263--279.

\bibitem[Schneider, 2004]{schneider_tropopause}
Schneider, T. (2004).
\newblock The tropopause and the thermal stratification in the extratropics of
  a dry atmosphere.
\newblock {\em Journal of the atmospheric sciences}, 61(12):1317--1340.

\bibitem[Schneider, 2006]{schneider_review}
Schneider, T. (2006).
\newblock The general circulation of the atmosphere.
\newblock {\em Annu. Rev. Earth Planet. Sci.}, 34:655--688.

\bibitem[Schwendike et~al., 2021]{schwendike}
Schwendike, J., Berry, G.~J., Fodor, K., and Reeder, M.~J. (2021).
\newblock On the relationship between the madden-julian oscillation and the
  hadley and walker circulations.
\newblock {\em Journal of Geophysical Research: Atmospheres},
  126(4):e2019JD032117.

\bibitem[Shaw and Voigt, 2015]{shaw2015tug}
Shaw, T. and Voigt, A. (2015).
\newblock Tug of war on summertime circulation between radiative forcing and
  sea surface warming.
\newblock {\em Nature Geoscience}, 8(7):560--566.

\bibitem[Shi and Bretherton, 2014]{shi2014large}
Shi, X. and Bretherton, C. (2014).
\newblock Large-scale character of an atmosphere in rotating
  radiative-convective equilibrium.
\newblock {\em Journal of Advances in Modeling Earth Systems}, 6(3):616--629.

\bibitem[Shi et~al., 2018]{shi2018wishe}
Shi, X. et~al. (2018).
\newblock Wishe-moisture mode in an aquaplanet simulation.
\newblock {\em Journal of advances in modeling earth systems},
  10(10):2393--2407.

\bibitem[Singh, 2022]{martin-notes}
Singh, M. (2022).
\newblock The general circulation of the atmosphere.

\bibitem[Singh and Kuang, 2016]{singhkuang2016}
Singh, M. and Kuang, Z. (2016).
\newblock Exploring the role of eddy momentum fluxes in determining the
  characteristics of the equinoctial hadley circulation: Fixed-sst simulations.
\newblock {\em Journal of the Atmospheric Sciences}, 73(6).

\bibitem[Singh et~al., 2017]{singhetal2017}
Singh, M., Kuang, Z., and Tian, Y. (2017).
\newblock Eddy influences on the strength of the hadley circulation: Dynamic
  and thermodynamic perspectives.
\newblock {\em Journal of the Atmospheric Sciences}, 74(2).

\bibitem[Suhas and Sukhatme, 2020]{suhas2020}
Suhas, D. and Sukhatme, J. (2020).
\newblock Moist shallow water response to tropical forcing: Initial value
  problems.
\newblock {\em Quarterly Journal of the Royal Meteorological Society}.

\bibitem[Suhas et~al., 2021]{suhas2021}
Suhas, D., Sukhatme, J., and Harnik, N. (2021).
\newblock Dry and moist atmospheric circulation with uniform sea-surface
  temperature.
\newblock {\em Quarterly Journal of the Royal Meteorological Society}.

\bibitem[Sukhatme, 2013]{sukhatme2013}
Sukhatme, J. (2013).
\newblock Longitudinal localization of tropical intraseasonal variability.
\newblock {\em Quarterly Journal of the Royal Meteorological Society},
  139(671):414--418.

\bibitem[Sukhatme, 2014]{sukhatme2014}
Sukhatme, J. (2014).
\newblock Low-frequency modes in an equatorial shallow-water model with
  moisture gradients.
\newblock {\em Quarterly Journal of the Royal Meteorological Society},
  140(683):1838--1846.

\bibitem[Sumi, 1992]{sumi}
Sumi, A. (1992).
\newblock Pattern formation of convective activity over the aqua-planet with
  globally uniform sea surface temperature (sst).
\newblock {\em Journal of the Meteorological Society of Japan. Ser. II},
  70(5):855--876.

\bibitem[Sun et~al., 2019]{sun2019regional}
Sun, Y., Li, L.~Z., Ramstein, G., Zhou, T., Tan, N., Kageyama, M., and Wang, S.
  (2019).
\newblock Regional meridional cells governing the interannual variability of
  the hadley circulation in boreal winter.
\newblock {\em Climate dynamics}, 52:831--853.

\bibitem[Tomassini and Yang, 2022]{tomassini2022}
Tomassini, L. and Yang, G.-Y. (2022).
\newblock Tropical moist convection as an important driver of atlantic hadley
  circulation variability.
\newblock {\em Quarterly Journal of the Royal Meteorological Society},
  148(748):3287--3302.

\bibitem[Toniazzo et~al., 2020]{toniazzo2020enforcing}
Toniazzo, T., Bentsen, M., Craig, C., Eaton, B.~E., Edwards, J., Goldhaber, S.,
  Jablonowski, C., and Lauritzen, P.~H. (2020).
\newblock Enforcing conservation of axial angular momentum in the atmospheric
  general circulation model cam6.
\newblock {\em Geoscientific Model Development}, 13(2):685--705.

\bibitem[Vallis, 2017]{vallis2017atmospheric}
Vallis, G.~K. (2017).
\newblock {\em Atmospheric and oceanic fluid dynamics}.
\newblock Cambridge University Press.

\bibitem[Vallis et~al., 2015]{vallis2015}
Vallis, G.~K., Zurita-Gotor, P., Cairns, C., and Kidston, J. (2015).
\newblock Response of the large-scale structure of the atmosphere to global
  warming.
\newblock {\em Quarterly Journal of the Royal Meteorological Society},
  141(690):1479--1501.

\bibitem[Voigt and Marotzke, 2010]{voigt_snowball}
Voigt, A. and Marotzke, J. (2010).
\newblock The transition from the present-day climate to a modern snowball
  earth.
\newblock {\em Climate dynamics}, 35:887--905.

\bibitem[Walker and Schneider, 2006]{walker2006eddy}
Walker, C. and Schneider, T. (2006).
\newblock Eddy influences on hadley circulations: Simulations with an idealized
  gcm.
\newblock {\em Journal of the atmospheric sciences}, 63(12):3333--3350.

\bibitem[Welch, 1967]{welch}
Welch, P. (1967).
\newblock The use of fast fourier transform for the estimation of power
  spectra: a method based on time averaging over short, modified periodograms.
\newblock {\em IEEE Transactions on audio and electroacoustics}, 15(2):70--73.

\bibitem[Wheeler and Kiladis, 1999]{wheeler1999convectively}
Wheeler, M. and Kiladis, G. (1999).
\newblock Convectively coupled equatorial waves: Analysis of clouds and
  temperature in the wavenumber--frequency domain.
\newblock {\em Journal of the Atmospheric Sciences}, 56(3):374--399.

\bibitem[Wheeler et~al., 2000]{WKW}
Wheeler, M., Kiladis, G., and Webster, P. (2000).
\newblock Large-scale dynamical fields associated with convectively coupled
  equatorial waves.
\newblock {\em Journal of the Atmospheric Sciences}, 57(5):613--640.

\bibitem[Williams, 1988a]{williamsA}
Williams, G.~P. (1988a).
\newblock The dynamical range of global circulations—i.
\newblock {\em Climate Dynamics}, 2(4):205--260.

\bibitem[Williams, 1988b]{williamsB}
Williams, G.~P. (1988b).
\newblock The dynamical range of global circulations—ii.
\newblock {\em Climate Dynamics}, 3:45--84.

\bibitem[Williamson et~al., 2013]{williamson2013aqua}
Williamson, D. et~al. (2013).
\newblock The aqua-planet experiment (ape): response to changed meridional sst
  profile.
\newblock {\em Journal of the Meteorological Society of Japan. Ser. II},
  91:57--89.

\bibitem[Williamson, 2008]{williamson2008equivalent}
Williamson, D.~L. (2008).
\newblock Equivalent finite volume and eulerian spectral transform horizontal
  resolutions established from aqua-planet simulations.
\newblock {\em Tellus A: Dynamic Meteorology and Oceanography}, 60(5):839--847.

\bibitem[Wu et~al., 2011]{wu_baroclinicity}
Wu, Y., Ting, M., Seager, R., Huang, H.-P., and Cane, M.~A. (2011).
\newblock Changes in storm tracks and energy transports in a warmer climate
  simulated by the gfdl cm2. 1 model.
\newblock {\em Climate dynamics}, 37:53--72.

\bibitem[Xie and Philander, 1994]{xie}
Xie, S.-P. and Philander, S. G.~H. (1994).
\newblock A coupled ocean-atmosphere model of relevance to the itcz in the
  eastern pacific.
\newblock {\em Tellus A}, 46(4):340--350.

\bibitem[Yuval and Kaspi, 2016]{yuval2016eddy}
Yuval, J. and Kaspi, Y. (2016).
\newblock Eddy activity sensitivity to changes in the vertical structure of
  baroclinicity.
\newblock {\em Journal of the Atmospheric Sciences}, 73(4):1709--1726.

\bibitem[Zaplotnik et~al., 2022]{zaplotnik}
Zaplotnik, {\v{Z}}., Pikovnik, M., and Boljka, L. (2022).
\newblock Recent hadley circulation strengthening: a trend or multidecadal
  variability?
\newblock {\em Journal of Climate}, 35(13):4157--4176.

\bibitem[Zhang, 2005]{zhang2005madden}
Zhang, C. (2005).
\newblock Madden-julian oscillation.
\newblock {\em Reviews of Geophysics}, 43(2).

\bibitem[Zhang and Wang, 2013]{zhang2013interannual}
Zhang, G. and Wang, Z. (2013).
\newblock Interannual variability of the atlantic hadley circulation in boreal
  summer and its impacts on tropical cyclone activity.
\newblock {\em Journal of Climate}, 26(21):8529--8544.

\bibitem[Zurita-Gotor, 2019]{zurita2019role}
Zurita-Gotor, P. (2019).
\newblock The role of the divergent circulation for large-scale eddy momentum
  transport in the tropics. part i: Observations.
\newblock {\em Journal of the Atmospheric Sciences}, 76(4):1125--1144.

\bibitem[Zurita-Gotor and {\'A}lvarez-Zapatero, 2018]{zurita2018coupled}
Zurita-Gotor, P. and {\'A}lvarez-Zapatero, P. (2018).
\newblock Coupled interannual variability of the hadley and ferrel cells.
\newblock {\em Journal of Climate}, 31(12):4757--4773.

\end{thebibliography}
\end{document}